\documentclass[12pt]{iopart}

\newcommand{\tabincell}[2]{\begin{tabular}{@{}#1@{}}#2\end{tabular}}
\usepackage{harvard}

\usepackage{graphicx}
\usepackage{amssymb}
\usepackage{tikz}
\usetikzlibrary{shapes.geometric, arrows}
\tikzstyle{squarered} = [rectangle, minimum width=3cm, minimum height=1cm, text centered, draw=black, fill=red!10]
\tikzstyle{squaregreen} = [rectangle, minimum width=3cm, minimum height=1cm, text centered, draw=black, fill=green!10]
\tikzstyle{squareblue} = [rectangle, minimum width=3cm, minimum height=1cm, text centered, draw=black, fill=blue!10]
\tikzstyle{diamondgreen} = [diamond, minimum width=3cm, minimum height=1cm, text centered, draw=black, fill=green!10]
\tikzstyle{diamondblue} = [diamond, minimum width=3cm, minimum height=1cm, text centered, draw=black, fill=blue!10]
\tikzstyle{diamondred} = [diamond, minimum width=3cm, minimum height=1cm, text centered, draw=black, fill=red!10]
\tikzstyle{diamondorange} = [diamond, minimum width=3cm, minimum height=1cm, text centered, draw=black, fill=orange!10]
\tikzstyle{circlewhite} = [circle, minimum width=0.5cm, minimum height=0.5cm, text centered, draw=black]
\tikzstyle{circleyellow} = [circle, minimum width=1cm, minimum height=1cm, text centered, draw=black, fill=yellow!10]
\tikzstyle{squareryellow} = [rectangle, rounded corners, minimum width=3cm, minimum height=1cm, text centered, draw=black, fill=yellow!10]
\tikzstyle{invisible} = [rectangle, minimum width=0.5cm, minimum height=0.5cm, text centered]
\tikzstyle{arrow} = [thick,->,>=stealth]
\tikzstyle{darrow} = [dashed,->,>=stealth]
\tikzstyle{line} = [thick,-,>=stealth]
\usepackage{makecell}
\usepackage{multirow}
\usepackage{rotating}
\usepackage{siunitx}
\usepackage[normalem]{ulem}
\usepackage{array,colortbl,xcolor}

\usepackage{booktabs} % the package for top/bottomrule

\def\aligned{\vcenter\bgroup\let\\\cr
\halign\bgroup&\hfil${}##{}$&${}##{}$\hfil\cr}
\def\endaligned{\crcr\egroup\egroup}

\usepackage{setspace}
\usepackage{lineno}
\usepackage{url}
\usepackage{soul}
\usepackage{xcolor}
\renewcommand{\hl}[1]{#1} %disable highlighting
\begin{document}

\title[Surrogate-free machine learning-based organ dose reconstruction]{Surrogate-free machine learning-based organ dose reconstruction for pediatric abdominal radiotherapy\footnote{This is the Accepted Manuscript version of an article accepted for publication in \PMB IOP Publishing Ltd is not responsible for any errors or omissions in this version of the manuscript or any version derived from it. This Accepted Manuscript is published under a CC BY licence. The Version of Record is available online at \url{https://doi.org/10.1088/1361-6560/ab9fcc}.}}

\author{M Virgolin\textsuperscript{1\footnote{shared first authorship, the two authors contributed equally to this work}}, Z Wang\textsuperscript{2\footnotemark[\value{footnote}]}, B V Balgobind\textsuperscript{2}, I W E M van Dijk\textsuperscript{2}, J  Wiersma\textsuperscript{2}, P S  Kroon\textsuperscript{3}, G O Janssens\textsuperscript{3,4}, M van Herk\textsuperscript{5}, D C Hodgson\textsuperscript{6}, L Zadravec Zaletel\textsuperscript{7}, C R N  Rasch\textsuperscript{8}, A Bel\textsuperscript{2}, P A N Bosman\textsuperscript{1,9}, T Alderliesten\textsuperscript{2,8}}

\address{$^1$ Life Sciences and Health Group, Centrum Wiskunde \& Informatica, the Netherlands\\
$^2$ Department of Radiation Oncology, Amsterdam UMC, University of Amsterdam, the Netherlands\\
$^3$ Department of Radiotherapy, University Medical Center Utrecht, the Netherlands\\
$^4$ Princess M\` axima Center for Pediatric Oncology, the Netherlands\\
$^5$ Manchester Cancer Research Centre, Division of Cancer Sciences, University of Manchester, United Kingdom\\
$^6$ Department of Radiation Oncology, Princess Margaret Cancer Centre, Canada\\
$^7$ Department of Radiation Oncology, Institute of Oncology Ljubljana, Slovenia\\
$^8$ Department of Radiation Oncology, Leiden University Medical Center, the Netherlands\\
$^9$ Department of Software Technology, Algorithmics Group, Delft University of Technology, the Netherlands\\
}

\ead{marco.virgolin@cwi.nl, z.wang@amsterdamumc.nl}
\vspace{10pt}
\begin{indented}

\item[] \textsuperscript{\footnotemark[\value{footnote}]} Shared first author, the two authors contributed equally to this work
%\item[]January 2020
\end{indented}

\newpage

\begin{abstract}
To study radiotherapy-related adverse effects, detailed dose information (3D distribution) is needed for accurate dose-effect modeling.
For childhood cancer survivors who underwent radiotherapy in the pre-CT era, only 2D radiographs were acquired, thus 3D dose distributions must be reconstructed from limited information. State-of-the-art methods achieve this by using 3D surrogate anatomies. These can however lack personalization and lead to coarse reconstructions.
We present and validate a surrogate-free dose reconstruction method based on Machine Learning (ML). 
Abdominal planning CTs ($n=142$) of recently-treated childhood cancer patients were gathered, their organs at risk were segmented, and $300$ artificial Wilms' tumor plans were sampled automatically. Each artificial plan was automatically emulated on the $142$ CTs, resulting in 42,600 3D dose distributions from which dose-volume metrics were derived.
Anatomical features were extracted from digitally reconstructed radiographs simulated from the CTs to resemble historical radiographs. Further, patient and radiotherapy plan features typically available from historical treatment records were collected. 
An evolutionary ML algorithm was then used to link features to dose-volume metrics. Besides 5-fold cross validation, a further evaluation was done on an independent dataset of five CTs each associated with two clinical plans.
Cross-validation resulted in mean absolute errors $\leq$0.6 Gy for organs completely inside or outside the field. For organs positioned at the edge of the field, mean absolute errors $\leq$1.7 Gy for D\textsubscript{mean}, $\leq$2.9 Gy for D\textsubscript{2cc}, and $\leq$13\% for V\textsubscript{5Gy} and V\textsubscript{10Gy}, were obtained, without systematic bias. Similar results were found for the independent dataset. To conclude, we proposed a novel organ dose reconstruction method that uses ML models to predict dose-volume metric values given patient and plan features. Our approach is not only accurate, but also efficient, as the setup of a surrogate is no longer needed.
%\noindent{\it Keywords\/}: dose reconstruction, radiotherapy dosimetry, machine learning, plan emulation, childhood cancer, late adverse effects
\end{abstract}

\vspace{2pc}
\submitto{\PMB}

(Some figures may appear in colour only in the online journal)
\maketitle

%\linenumbers

\section{Introduction}
\label{sec:introduction}

Patients undergoing radiotherapy (RT) are prone to develop radiation-related Adverse Effects (AEs) \cite{birgisson2005adverse,van2010evaluation,cheung2017chronic}. 
To improve the design of future multi-modality treatments, clinicians are interested in better understanding the relationship between radiation dose and onset of AEs.
Modern research efforts in this direction delve into dosimetric details, employing dose distribution metrics to a specific organ (or sub-volume) as explanatory variables. Such rich information is obtained by simulating the RT plan on 3D imaging of the patient (i.e., CT scans) with organ segmentations in a Treatment Planning System (TPS) \cite{donovan2007randomised,feng2007intensity,bolling2011dose}. 

Unfortunately, when so-called \emph{late} AEs (onset can be decades after RT) need to be studied, it is not always possible to straightforwardly obtain detailed information on dose distributions \cite{birgisson2005adverse}. For patients who underwent RT before the use of planning CTs became commonplace (in the following, \emph{historical patients}), 2D radiographs were used for treatment planning (e.g., this was the case until the 1990s in the Netherlands \cite{van2010evaluation}), meaning no 3D anatomical imaging is available. Consequently, no simulations can be performed in a TPS to estimate 3D dose distributions for these patients \cite{stovall2006dose,verellen2008short,ng2012individualized}. The information available for historical patients normally consists of what was reported in treatment records, e.g., features of the patient such as age and gender, and features of the plan such as prescribed dose, geometry of the plan, and the use of blocks. Additionally, 2D radiographs can be available, from which information can be gathered on the internal anatomy (mainly bony anatomy, as internal organs are normally not clearly distinguishable), and on the plan configuration with respect to the patient's anatomy \cite{leisenring2009pediatric,van2010evaluation}.

To improve the understanding of late AEs, recent research is striving to develop increasingly accurate \emph{dose reconstruction} methods, i.e., methods to estimate the 3D dose distribution received by historical patients \cite{stovall2006dose,ng2012individualized,xu2014exponential,lee2015reconstruction}. State-of-the-art approaches employ \emph{phantoms}, i.e., 3D surrogates of the human anatomy upon which the RT plan can be simulated, to compute the dose distribution. Phantoms exist in different forms: physical or virtual, made by simple geometrical shapes or by adopting and morphing actual CT scans and organ segmentations \cite{stovall2006dose,xu2014exponential,lee2015reconstruction}. Generally, phantoms are built to represent \emph{average} anatomies, for categories of patients (e.g., for a certain age range), and are collected into so-called phantom libraries \cite{cassola2011standing,segars2013XCATlibrary,geyer2014uf}. Whenever dose reconstruction for a historical patient is needed, the phantom that represents the category that the patient belongs to is retrieved from the library and used as surrogate for simulation of the RT plan. 

As the largest source of error related to phantom-based dose reconstruction comes from the mismatch between the anatomy of the phantom and the true anatomy of the patient \cite{bezin2017review}, it is important to define the best way to match phantoms to patients. This issue is still under research, and different approaches employ different heuristic matching criteria that are normally hand-crafted and based upon statistics and guidelines drawn from large population studies (e.g., ICRP89, NANTHES) \cite{icrp89,cassola2011standing,segars2013XCATlibrary,geyer2014uf}. However, the use of heuristic matching criteria has been hypothesized to be too simplistic to capture the high variability of internal human anatomy \cite{de2001organ,geyer2014uf,xu2014exponential,virgolin2018feasibility,wang2019how}. For example, a popular phantom-based dose reconstruction approach uses solely age and gender for surrogate matching \cite{howell2019adaptations}. Our group's recent work focusing on Wilms' tumor (the most common type of kidney cancer for childhood cancer patients) irradiation for pediatric patients showed that utilizing surrogate CTs using age- and gender-based matching can lead to poor dose reconstruction quality in individual cases \cite{wang2018age}.

To improve the resemblance of a surrogate phantom, there have been efforts to replace the normally hand-crafted heuristic matching criteria with data-driven decisions. For example, statistical models inferred from CTs and 3D organ segmentations of adult patients have been used to drive a deformable image registration procedure that adapted 3D organ segmentations to the 2D anatomy of a specific patient, given features of the latter as measurable from 2D radiographs \cite{ng2012individualized,mishra2013evaluation}. Using a state-of-the-art Machine Learning (ML) algorithm, it has been shown that features typically available for historical patients treated for Wilms' tumor can be linked to different 3D anatomy similarity metrics based on organ segmentations and CTs \cite{virgolin2018feasibility}. Our group recently proposed an automatic pipeline that uses ML to steer the assembling of a new original anatomy based on 3D CTs and organ segmentations of multiple patients using the features of a historical patient \cite{virgolin2019machine,virgolin2020spie}. However, it is important to realize that maximizing some form of overall anatomical resemblance is difficult. Moreover, from the standpoint of optimizing dose reconstruction accuracy, it can be considered sub-optimal for RT dosimetry purposes. This is because in RT dosimetry, what part of anatomy is most meaningful largely depends on the particular RT plan \cite{wang2019how}. 

To the best of our knowledge, although \emph{both} patient anatomy and plan geometry play a key role in determining dose-volume metrics for Organs At Risk (OARs), existing dose reconstruction approaches focused solely on patient anatomy information, to obtain a representative surrogate. Plan information is used only later, to calculate the dose on the surrogate. The purpose of this article is to develop and validate an ML approach to predict dose-volume metrics for OARs based on patient anatomy and plan geometry information. Specifically, we propose to use ML to directly learn what dose-volume metrics for an OAR are likely given information on the patient and on the plan, without the need to select or craft any surrogate anatomy. We argue that this is a sensible choice because ML can directly be trained upon what ultimately matters, i.e., dose reconstruction accuracy. 
Further, we present a method to generate artificial plans automatically, to obtain sufficient data to train and validate our ML-based dose reconstruction approach. In addition, we assess whether training ML on artificial plans generalizes to a small set of clinical plans.

\section{Materials \& Methods}

We considered pediatric flank RT, and in particular RT for Wilms' tumor, as an application for our dose reconstruction method, in continuity with our previous work. The choice to focus on pediatrics is because children are the most prone to develop late AEs \cite{cheung2017chronic}, and are typically underrepresented in existing phantom libraries \cite{xu2014exponential}. Moreover, more than 85\% of pediatric patients survives Wilms' tumor five years or longer, but considerable chances of the onset of late AEs remain \cite{van2010evaluation}.

\subsection{Patient data}\label{sec:recentpatientdata}

To be able to create a ground-truth to learn dose-volume metrics from, CT scans were needed. Hence, a total of 142 pediatric planning CTs were collected by involving the following institutes (number of CTs in brackets): Amsterdam University Medical Centers / Emma Children's Hospital ($n=38$), University Medical Center Utrecht / Princess M\'axima Center for Pediatric Oncology ($n=42$), The Christie NHS Foundation Trust ($n=33$), Princess Margaret Cancer Centre ($n=18$), and Institute of Oncology Ljubljana ($n=11$). Five further CTs were collected from the Amsterdam University Medical Centers and kept aside to be used exclusively for an additional assessment (Sec.~\ref{sec:method_clinical_validation}).

The inclusion criteria were: patient age at scan acquisition between 1 to 8 years; the CT field of view including a common abdominal region from the tenth thoracic (T10) vertebral body to the first sacral (S1) vertebral body; presence of five lumbar vertebrae (rare cases of patients with six exist); patient scanned in supine position; quality of CT sufficient to perform organ segmentations. The patients underwent RT between 2002 and 2018, mostly but not exclusively for abdominal cancers. The median CT slice resolution was 
$0.94\times0.94$ mm, the median slice thickness was $3$ mm.

As we focused on Wilms' tumor treatment, four OARs were considered: the liver, the spleen, the contralateral kidney (left or right, depending on the side of the tumor), and the spinal cord (between T10 and S1). We prepared the OAR segmentation in all CTs ($n=142+5$) by either manual or automatic segmentation, followed by potential adaptation and approval by two clinical experts (I.W.E.M. van Dijk and B.V. Balgobind). 
Some patients did not have both kidneys intact, due to nephrectomy prior to RT. The number of CTs that had only a complete right kidney, only a complete left kidney, and two complete kidneys were 36, 40, and 71, respectively.

\subsection{Automatic generation of artificial Wilms' tumor plans}\label{sec:artificial-plan-generation}
A method to automatically generate historical-like abdominal flank irradiation plans (i.e., artificial plans) for Wilms' tumor treatment based on information visible on 2D radiographs was created, in order to obtain large plan variations. 

\begin{figure}
    \centering
    \includegraphics[width=1.0\linewidth]{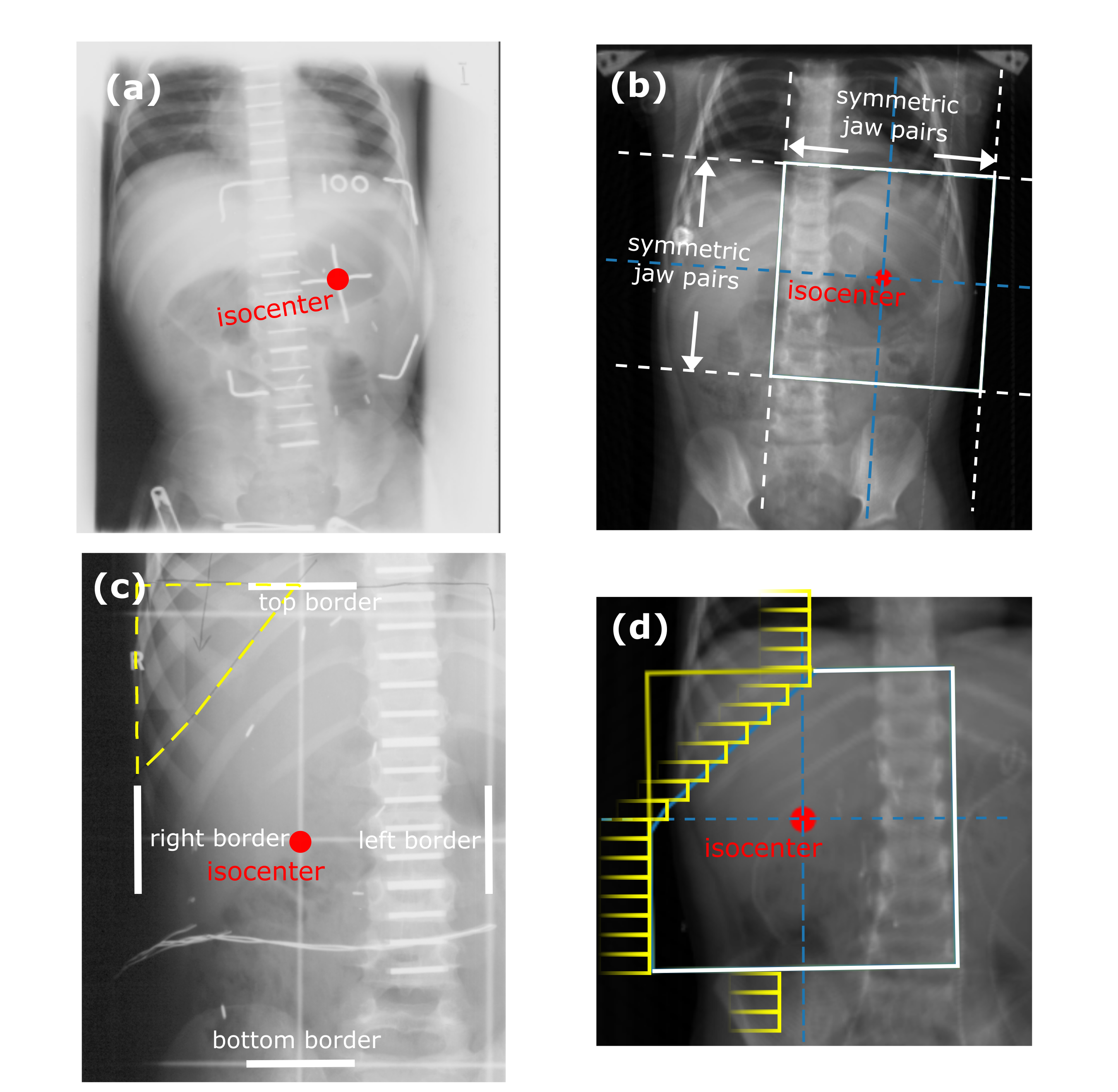}
    \caption{(a) An actual hand-drawn plan on a historical radiograph with a rectangular field (indicated by white corners). (b) An artificial plan with a rectangular field (in white lines) plotted on the DRR of a recent patient. (c) An actual hand-drawn plan on a historical radiograph with a rectangular field (in white bars) and an additional block (outlined by dashed yellow lines) to spare part of the liver. (d) An artificial plan plotted on the DRR of a recent patient with a rectangular field (in white bars) and an additional block (obtained by multi-leaf collimators, outlined by yellow lines) to spare part of the liver. For each plot, the isocenter is indicated by a red dot in the middle of the field.}
    \label{fig:historicalplans}
\end{figure}

Figures~\ref{fig:historicalplans}(a) and \ref{fig:historicalplans}(c) illustrate examples of actual historical plans on respective historical radiographs. As can be observed, a typical historical flank irradiation field is a rectangular area, with possible shielding blocks, that is located on the right or on the left flank. Irradiation is done by beams from anterior-posterior (AP) and posterior-anterior (PA) direction. Along right-left (RL), one field border is located at the edge of the patient's body contour, while the other is located as to include the vertebral column \cite{SIOPprotocol}. In some cases, blocks are placed to protect OARs from irradiation (Fig.~\ref{fig:historicalplans}(c)). In historical plans the isocenter is positioned in the center of the treatment field that is projected on the coronal plane (Fig.~\ref{fig:historicalplans}) and at the middle of the patient's AP abdominal diameter. 

\begin{figure}
    \centering
    \includegraphics[width=0.7\linewidth]{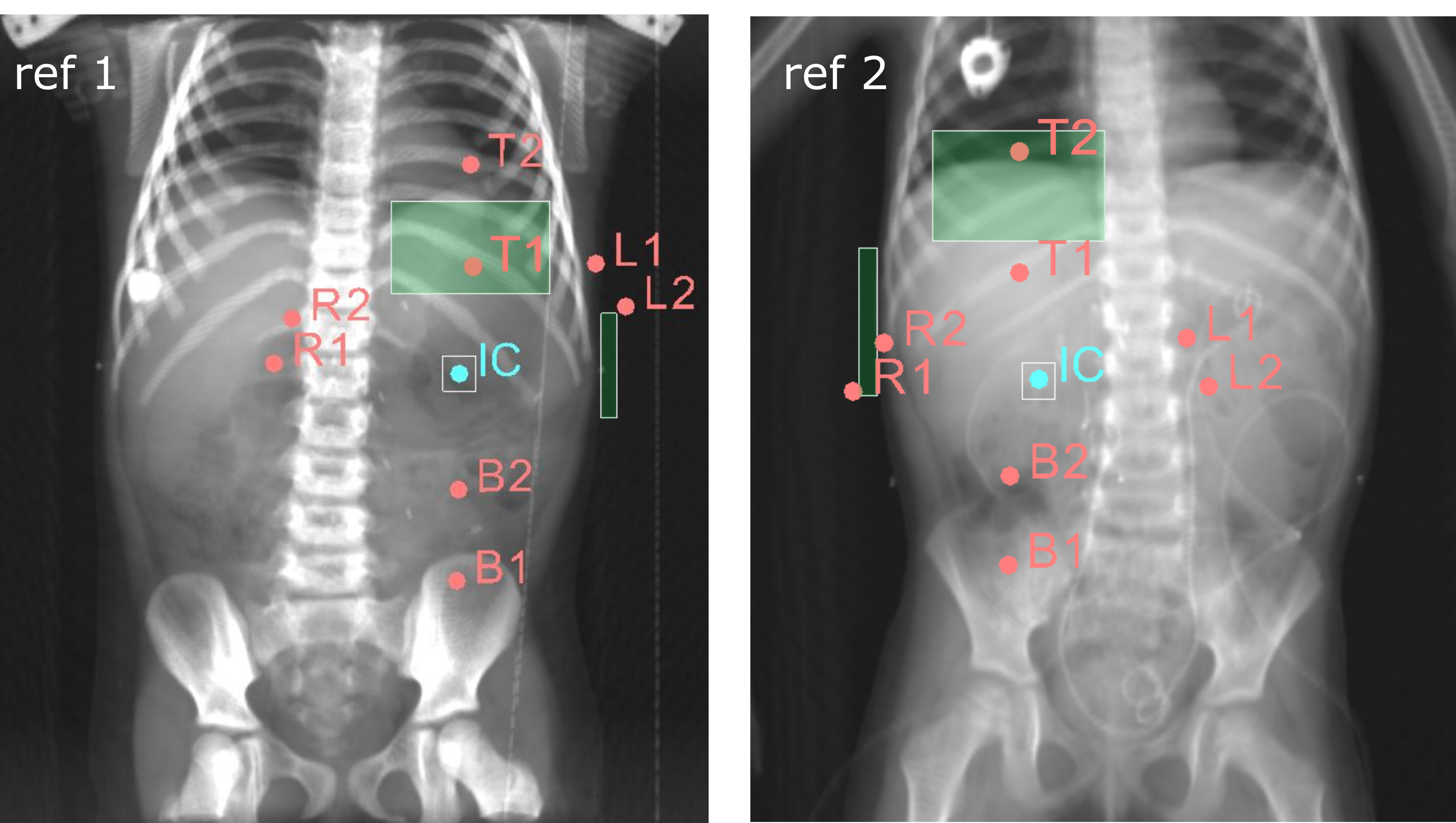}
    \caption{Examples of landmark locations to specify geometry variability of two types of artificial plans (left-sided plans in the left figure and right-sided plans in the right figure). Ref 1 is the DRR derived from the reference CT of a 5-year-old female patient and ref 2 is the DRR derived from the reference CT of a 4-year-old female patient. The box around the isocenter (IC) specifies the range of possible isocenter positions. The vertical position of T1/T2 and of B1/B2 specify the lowest/highest position of the upper and lower border of the field, respectively. The horizontal positions of R1/R2 and L1/L2 specify the rightmost/leftmost position of the right and left border of the field, respectively. The isocenter and artificial field border positions were sampled uniformly at random within the specified ranges. The green boxes indicate the regions where two endpoints of a line representing a block border can be sampled. This line, together with the upper and left/right field borders, encloses the block.}
    \label{fig:artificial}
\end{figure}

To generate artificial plans, two reference digitally reconstructed radiographs (DRRs) were considered, randomly selected from the data. One DRR was derived from a CT of a 5-year old female patient without nephrectomy (ref 1 in Fig.~\ref{fig:artificial}), and the other was derived from a CT of a 4-year old female patient with nephrectomy of the left kidney (ref 2 in Fig.~\ref{fig:artificial}).
Upon these two DRRs, boundaries defining the extent of variation for clinically reasonable fields were identified by an experienced pediatric radiation oncologist (B. V. Balgobind)
Note that historical clinical guidelines are slightly different from current ones (e.g., currently the iliac crests should be safeguarded, unlike in Fig.~\ref{fig:historicalplans}(c)).
Figure~\ref{fig:artificial} shows two examples of landmark locations identifying possible plan variations, on the two reference DRRs. 
Specifically, given the boundaries of possible isocenter positions and field borders, plans with a rectangular field were generated by sampling \emph{uniformly} within those boundaries. 

For each plan generated, an additional version of that plan including one block was generated as well. A block was simulated as the area in the upper lateral corner enclosed by the border of the rectangular field and a line crossing two randomly sampled endpoints. The endpoints were sampled from two regions roughly covering the start and end points of rib 9 and rib 12 on the DRRs (regions indicated by the green boxes in Fig.~\ref{fig:artificial}). This way, a sampled block covered part of the liver (in right-sided plans) or part of the spleen (in left-sided plans). All plans consisted of two opposing and symmetrical beams in AP-PA directions irradiating one side of the abdominal flank. Figures~\ref{fig:historicalplans}(b) and \ref{fig:historicalplans}(d) illustrate two examples of sampled artificial plans (without or with a block) on respective DRRs.

A total of 300 artificial plans were generated automatically, of which 150 without a block, and 150 with a block. 
The random sampling of the plan side led to 142 left-sided plans and 158 right-sided plans.
The same set of plan features used in our previous work was considered to generate plans in DICOM RTPLAN format (e.g., gantry and collimator angles, isocenter location, field sizes) \cite{wang2020spiej}.

\subsection{Generation of the dataset for ML}\label{sec:generation-ML-dataset}
\begin{figure}
    \centering
    \includegraphics[width=1.0\linewidth]{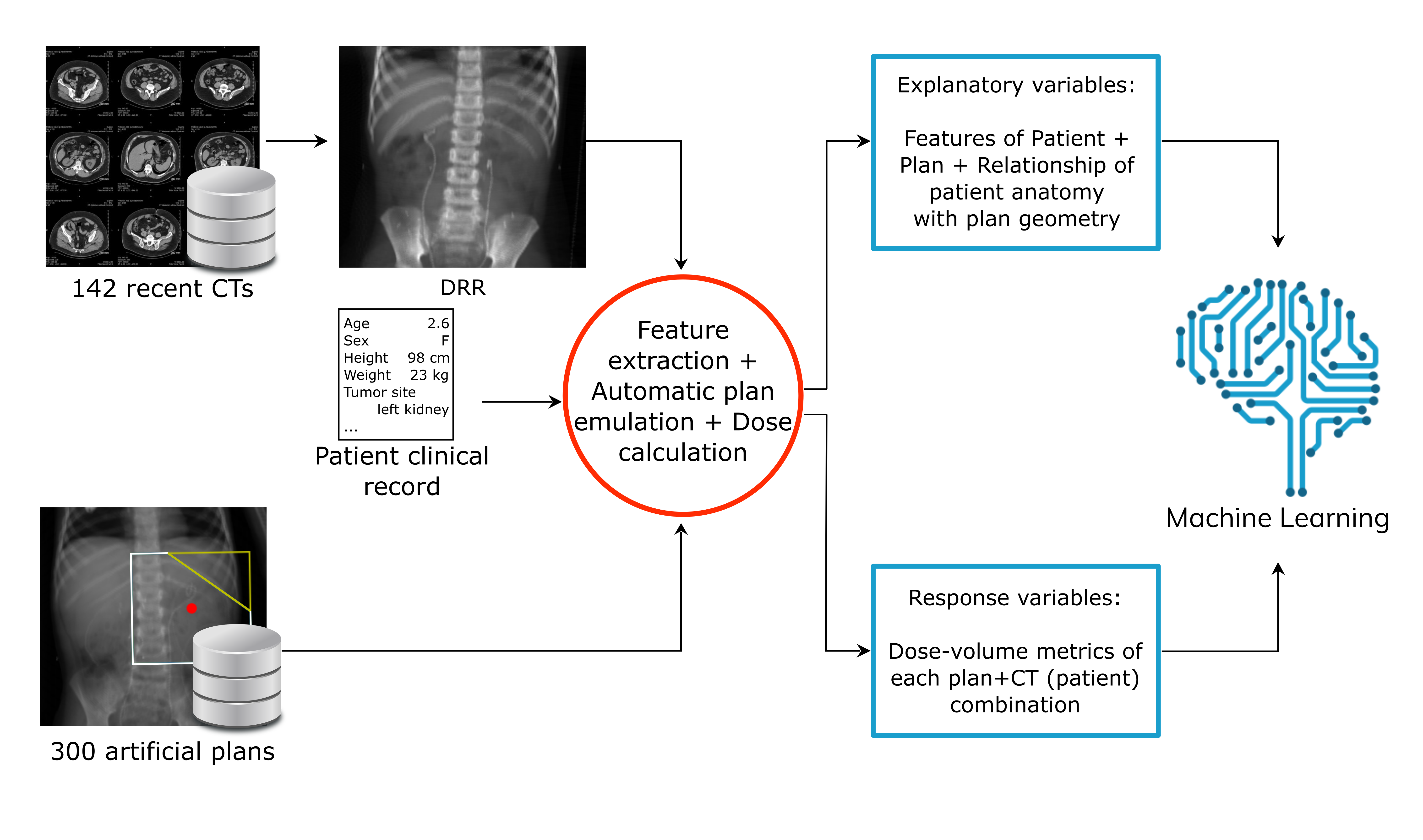}
    \caption{Pipeline for data generation. Artificial plans are sampled automatically. The explanatory and response variables are used as input to train the ML model. The explanatory variables include features of the plan (e.g., isocenter location, field size), patient features (e.g., age, nephrectomy), and features on the relationship between the anatomy of the patient and the geometry of the plan (e.g., signed distance between the 2\textsuperscript{nd} lumbar vertebra and the plan isocenter). The response variables are dose-volume metrics for each OAR.}
    \label{fig:datagenerationML}
\end{figure}

Figure~\ref{fig:datagenerationML} summarizes the pipeline used to generate the dataset for ML. Firstly, we emulated each of the 300 artificial plans on each of the 142 CT scans by the automatic plan emulation method proposed in our previous work \cite{wang2020spiej}, leading to a total of 42,600 emulations. 
The method automatically transfers a plan prepared on one CT to another CT (with quality comparable to human experts), using landmark detection upon the respective DRRs.
Secondly, for each of the 42,600 plan emulations, dose-volume metrics of interest (see Sec.~\ref{sec:planemulation}) were collected for the different OARs by use of our automatic dose computation pipeline \cite{wang2020spiej}. The pipeline used the collapsed cone dose calculation algorithm of Oncentra TPS (version 4.3, Elekta AB, Stockhom, Sweden). Thirdly, features that are plausible to be available for typical historical cases were collected from the anatomy of the included CTs as visible in the respective DRRs, from the artificial plans, and from the relationship between anatomy and plan geometry.

\subsubsection{Response variables: dose-volume metrics}\label{sec:planemulation}
To select dose-volume metrics to use as response variables for ML, we considered metrics typically used to validate state-of-the-art dose reconstruction approaches \cite{ng2012individualized,lee2015reconstruction}, and typically found to be of clinical relevance in studies of AEs in adults (e.g., QUANTEC) \cite{leisenring2009pediatric,bolling2011dose,emami2013tolerance}. Studies on dose-volume response relationships for pediatric patients (so-called PENTEC studies \cite{constine2019pediatric}) are currently limited.
We considered mean organ dose ($D\textsubscript{mean}$), the minimum dose received by the maximally exposed 2 cubic centimeters of an OAR ($D\textsubscript{2cc}$) (similar but more robust than the maximum dose to a single point), the percentage of OAR volume receiving at least $5$ Gy ($V\textsubscript{5Gy}$), and the percentage of OAR volume receiving at least $10$ Gy ($V\textsubscript{10Gy}$). 
Regarding $D\textsubscript{2cc}$, we included this metric because peak dose values to a small OAR portion may be relevant to explain late AEs related to OARs that work in a serial fashion (e.g., the spinal cord).

\subsubsection{Explanatory variables: features of patients and plans}\label{sec:features-used}
To assess what information can be available for historical patients, we considered the Dutch records of the Emma Children's Hospital/Academic Medical Center childhood cancer survivor cohort, who underwent RT between 1966 and 1996 \cite{van2010evaluation}. For this cohort, along with historical patient records and treatment plan details, 2D coronal radiographs were consistently taken, hence providing partial information on the anatomy.

The complete set of features considered in this work is reported in Table~\ref{tab:historical-features}. Note the absence of height and weight (which are used by some phantom-based methods \cite{geyer2014uf}). For 12\% of the patients, height and/or weight data were missing and preliminary experiments using automatic imputation methods showed no benefit in including them. 

For the features related to anatomical geometry, anatomical landmarks from DRRs were detected automatically using the landmark detection method in our previous work \cite{wang2020spiej}. Note that the landmarks concerned only bony anatomy because other internal anatomy tissues are not reliably visible in historical radiographs. Importantly, we normalized features related to measurements of anatomy and anatomy-plan geometry configuration (e.g., rib-cage width, field sizes, distances between landmarks and the isocenter) by the width and height of the respective DRR they were measured from (after the DRRs were cropped to a same region of interest between T10 and S1). This was done because when plans are emulated, they are scaled based on proportions derived from the landmarks \cite{wang2018age,wang2020spiej}. Since differences in anatomy solely due to overall anatomy scaling do not result in different dose-volume metric values, these differences should not be accounted for by the explanatory variables (confirmed in preliminary experiments).

The abdominal diameter in AP ($\textit{Diam}^\textit{IC}_\textit{AP}$) is the only anatomical feature not measurable from DRRs generated along AP/PA direction. In historical RT, it was measured using a ruler to determine the isocenter position along the AP axis, and was subsequently reported in the records. 
For our cohort, we measured $\textit{Diam}^\textit{IC}_\textit{AP}$ automatically on the CT scans, by using a pre-determined isocenter position of typical abdominal flank irradiation plans. In particular, the isocenter position along the inferior-superior (IS) axis was set to the intervertebral disk between the 1\textsuperscript{st} and the 2\textsuperscript{nd} lumbar vertebra (L1 and L2), while the center of mass of the kidney was used to determine the isocenter position along RL (for CTs including both kidneys, $\textit{Diam}^\textit{IC}_\textit{AP}$ was measured twice and the  average was taken).

In our simulations we used for all artificial plans the same fractionation scheme (8 $\times$ 1.8 Gy), beam energy (6 MV), and prescribed dose (14.4 Gy at isocenter). 
These settings are the most common in historical records, and are still valid in the current Wilms' tumor RT protocol \cite{SIOPprotocol}. 
Moreover, choosing a specific prescribed dose (e.g., 14.4 Gy) does not limit generalizability, since the dose distribution over the entire anatomy depends linearly on the prescribed dose (i.e., it can be rescaled).

\begin{figure}
    \centering
    \includegraphics[width=1.0\linewidth]{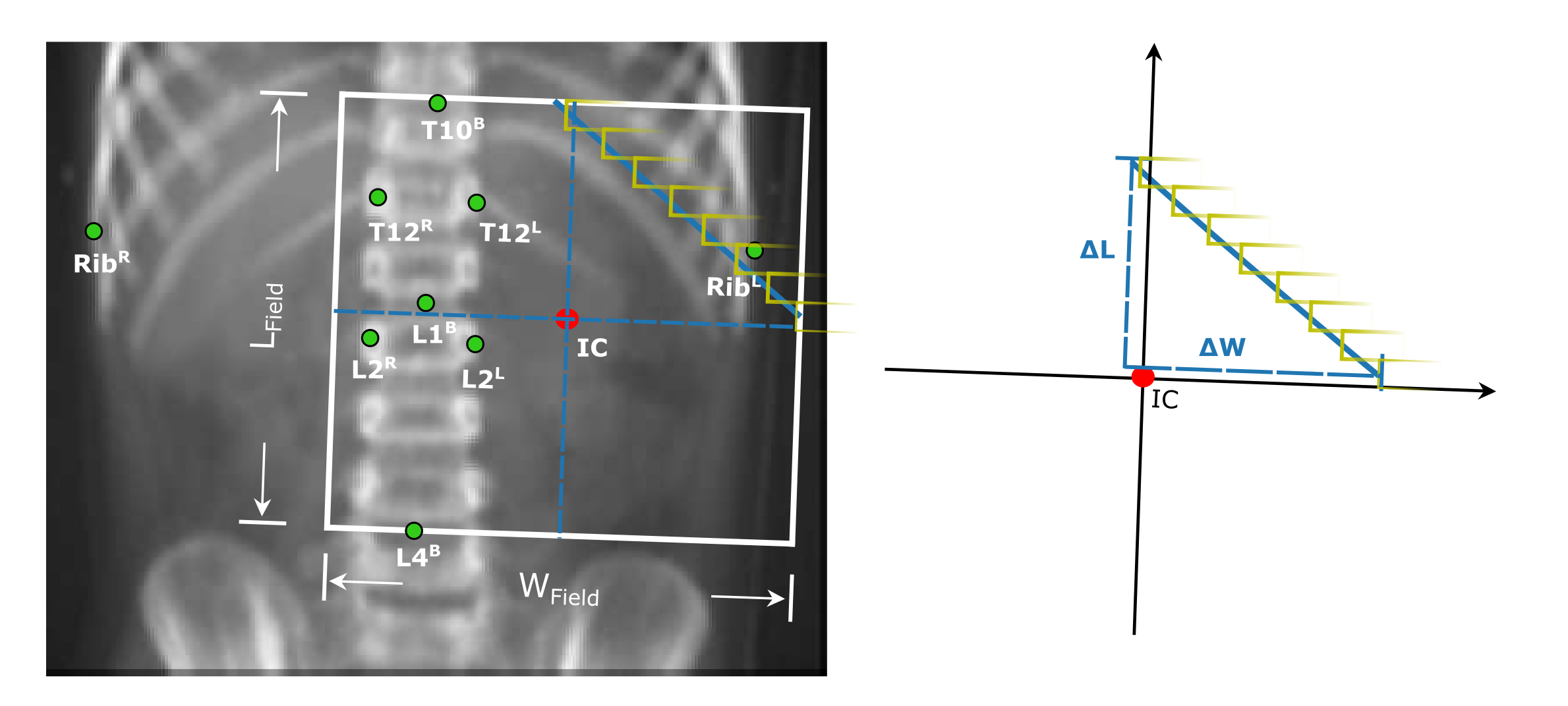}
    \caption{An example of the beam's eye view of a plan plotted on a DRR with the landmark locations used to compute the features concerning plan field configuration on top of the patient's anatomy. The plot next to the DRR illustrates how the block is simulated by aligning the center of the leaves with the boundary of the block and how the slope of an MLC-simulated block is calculated.}
    \label{fig:landmarks}
\end{figure}

For both fields with and without a block, the features representing field sizes in RL and IS directions ($W_{\textit{Field}}$ and $L_{\textit{Field}}$) were set by simply considering the full rectangular area (i.e., irrespective of blocking). For fields with a block, the slope of the block (note that the block is formed by Multi-Leaf Collimators (MLCs)) and the ratio between the blocked region and non-blocked region of the field ($\textit{Ratio}_\textit{\scriptsize Block}$) was computed. In addition, we considered features that relate to how the plan was configured with respect to the patient's anatomy, based on the position of the isocenter and of the bony landmarks. For instance, $\Delta^\textit{IC}_\textit{RL}(T10^B)$ links the bottom of the T10 vertebra to the position of the isocenter in RL direction. Figure~\ref{fig:landmarks} shows an example of the anatomical landmarks and plan geometrical borders used to calculate features describing plan configuration with respect to the patient's anatomy. 

\begin{table}[]
     \fontsize{8pt}{11pt}\selectfont
     \renewcommand{\arraystretch}{1.6} % expand the row height of the table
    \caption{Description of the 33 features considered as explanatory variables for ML. }
    \smallskip
    \label{tab:historical-features}
    \centering
    \scriptsize
    \scalebox{0.85}{
    \begin{tabular}{lccl}
    \toprule
        \textbf{Feature name} & \textbf{Origin} & \textbf{Unit} & \textbf{Description} \\ 
        \hline
        \textit{Age} & Records & years & patient age at CT scanning\\
        \textit{ArmsUp} & DRR & yes/no & whether the patient had arms in a raised position during scanning \\
        $\textit{Diam}^\textit{IC}_\textit{AP}$ & Records & cm & patient AP diameter measured at isocenter \\
        \textit{Nephrectomy} & Records & yes\slash no & whether the patient underwent nephrectomy \\
        $W_{\textit{Rib}^R}$  & DRR & cm & width (in RL) of right-part of the rib cage (from vertebral column to location of right-most rib)\\
        $W_{\textit{Rib}^L}$ & DRR & cm & width (in RL) of left-part of the rib cage (from vertebral column to location of left-most rib) \\
        $W_\textit{VC}$ & DRR & cm & average vertebral column width\\
        $L_\textit{VC}$ & DRR & cm & length (in IS) of the vertebral column from T11 to L4\\
        $W_{\textit{Field}}$ & Plan & cm & field width (in RL) \\
        $L_{\textit{Field}}$ & Plan & cm & field length (in IS) \\
        \textit{FieldSide} & Plan & right\slash left & whether the plan concerns left-sided or right-sided flank irradiation \\
        $\textit{Intercept}_\textit{\scriptsize Block}$ & Plan & cm & distance (in RL) between isocenter and block endpoint of the top field border \\
        $\textit{Ratio}_\textit{\scriptsize Block}$ & Plan & \% & $\textit{Area}(\textit{Block})/\textit{Area}(\textit{Rectangular field})$, 0 for block-free plans \\
        $\textit{Slope}_\textit{\scriptsize Block}$ & Plan & - & $\Delta L / \Delta W$ of the block (see Fig.~\ref{fig:landmarks}); 0 for block-free plans \\
        $\theta_{C}$ & Plan & $\circ$ & angle of collimator system with respect to gantry system \\
        $\Delta^\textit{IC}_\textit{RL}(T10^B)$ & Plan $+$ DRR & cm & RL distance between bottom of T10 and isocenter \\
        $\Delta^\textit{IC}_\textit{IS}(T10^B)$ & Plan $+$ DRR & cm & IS distance between bottom of T10 and isocenter \\
        $\Delta^\textit{IC}_\textit{RL}(T12^R)$ & plan $+$ DRR & cm & RL distance between right border of T12 and isocenter \\
        $\Delta^\textit{IC}_\textit{IS}(T12^R)$ & Plan $+$ DRR & cm &  IS distance between right border of T12 and isocenter \\
        $\Delta^\textit{IC}_\textit{RL}(T12^L)$ & Plan $+$ DRR & cm &  RL distance between left border of T12 and isocenter \\
        $\Delta^\textit{IC}_\textit{IS}(T12^L)$ & Plan $+$ DRR & cm &  IS distance between left border of T12 and isocenter \\
        $\Delta^\textit{IC}_\textit{RL}(L1^B)$ & Plan $+$ DRR & cm &  RL distance between bottom of L1 and isocenter \\
        $\Delta^\textit{IC}_\textit{IS}(L1^B)$ & Plan $+$ DRR & cm &  IS distance between bottom of L1 and isocenter \\
        $\Delta^\textit{IC}_\textit{RL}(L2^R)$ & Plan $+$ DRR & cm &  RL distance between right border of L2 and isocenter \\
        $\Delta^\textit{IC}_\textit{IS}(L2^R)$ & Plan $+$ DRR & cm &  IS distance between right border of L2 and isocenter \\
        $\Delta^\textit{IC}_\textit{RL}(L2^L)$ & Plan $+$ DRR & cm &  RL distance between left border of L2 and isocenter \\
        $\Delta^\textit{IC}_\textit{IS}(L2^L)$ & Plan $+$ DRR & cm &  IS distance between left border of L2 and isocenter \\
        $\Delta^\textit{IC}_\textit{RL}(L4^B)$ & Plan $+$ DRR & cm &  RL distance between bottom of L4 and isocenter \\
        $\Delta^\textit{IC}_\textit{IS}(L4^B)$ & Plan $+$ DRR & cm &  IS distance between bottom of L4 and isocenter \\
        $\Delta^\textit{IC}_\textit{RL}(\textit{Rib}^R)$ & Plan $+$ DRR & cm & RL distance between location of right-most rib and isocenter \\
        $\Delta^\textit{IC}_\textit{IS}(\textit{Rib}^R)$ & Plan $+$ DRR & cm & IS distance between location of right-most rib and isocenter \\
        $\Delta^\textit{IC}_\textit{RL}(\textit{Rib}^L)$ & Plan $+$ DRR & cm & RL distance between location of left-most rib and isocenter \\
        $\Delta^\textit{IC}_\textit{IS}(\textit{Rib}^L)$  & Plan $+$ DRR & cm & IS distance between location of left-most rib and isocenter \\
        \bottomrule
    \end{tabular}
    }
    \raggedright
    Abbreviations: R (in superscript): right, L (in superscript): left, RL: right-left, AP: anterior-posterior, IS: inferior-superior, IC: isocenter, VC: vertebral column, W: width, L in $L_\textit{VC}$ and $L_{\textit{Field}}$ : length.
\end{table}

\subsubsection{Dataset for supervised learning}
Features and dose-volume metrics were finally collected in a dataset. The dataset corresponded to a 2D matrix, where the rows represented patient-plan combinations, i.e., examples ($n=42,600$), and the columns represented features ($33$) and response variables ($4$ for each OAR).

\subsection{Machine learning}\label{sec:machinelearning}
In the following sections we describe how ML was performed in terms of training and validation on the artificial plans. We further introduce the ML algorithms adopted and a control method, and describe the assessment on clinical plans aimed at evaluating whether artificial plans are good representatives of clinical plans to train ML models.

\subsubsection{Training and evaluation of ML models}\label{sec:training-and-eval-ml}
Since dose metrics are scalars, we treated the learning problem as a regression problem. We trained a separate ML model for each combination of dose-volume metric and OAR.

Preliminary analysis showed that right-sided plans and left-sided plans led to markedly different distributions of possible dose-volume metric values for all OARs except for the spinal cord. Thus, ML models were set to be composed of two sub-models, each to be trained independently on a particular sub-set of the data based on plan side (right or left).

The quality of the models was estimated with a 5-fold cross-validation. This means that a random partition of 1/5th of the total number of patients and plans was held out (test set), and training was performed on the remaining data. Then, the prediction error was measured on the test set. This process was repeated five times, each time considering a different data partition for the test set. No patient nor plan that was in the test set was included in the data at training time.

Each training step included hyper-parameter tuning by grid-search with internal 5-fold cross-validation (upon the training set), as well as feature selection (which resulted in eight features being systematically discarded, see the supplementary material A).
For each dose-volume metric $k \in \{ D_\textsubscript{mean}, D_\textsubscript{2cc}, V_\textsubscript{5Gy}, V_\textsubscript{10Gy} \}$, the Root Mean Square Error (RMSE) loss was used, i.e., 
\begin{equation}
    \textit{RMSE} ( Y^k, \hat{Y}^k) = \sqrt{ \frac{1}{\nu} \sum_{i=1}^\nu \left( Y^k_i -  \hat{Y}^k_i \right)^2 },
\end{equation}
where $Y^k$ are the ground truth values and $\hat{Y}^k$ are the model predictions for the dose-volume metric $k$, and $\nu$ is the total number of rows in the training set. The RMSE was chosen to regularize ML, i.e., to penalize larger errors more \cite{bishop2006pattern}.

To account for the stochastic nature of one of the ML algorithms employed (see Sec.~\ref{sec:gpgomea}), and the random partitioning of the data, the 5-fold cross-validation was repeated ten times. The averages and standard deviations over the $5\times10$ validation results (five folds repeated ten times) were considered.

\subsubsection{Machine learning algorithms and control method}\label{sec:gpgomea}
We considered two ML algorithms. The first one was the ELastic Net (ELN) \cite{zou2005regularization}, a very popular baseline for regression which combines regularization of the ridge and lasso methods \cite{hoerl1970ridge,tibshirani1996regression}.
The second ML algorithm we considered was the recently introduced Genetic Programming version of the Gene-pool Optimal Mixing Evolutionary Algorithm (GP-GOMEA), as it was found to achieve competitive performance on a variety of benchmark problems (including regression ones \cite{virgolin2017scalable,virgolin2019model}, as well as in previous work concerning radiotherapy \cite{virgolin2018symbolic,virgolin2019machine}.
Details on the hyper-parameters of ELN and GP-GOMEA and on their tuning are reported in the supplementary material B.

We further considered a control method inspired by phantom-based dose reconstruction as performed by~\cite{howell2019adaptations}, where a phantom is chosen as representative surrogate based on age and gender similarity, and the dose is calculated by emulating the plan on the phantom. In our case, given a test patient, we selected the CT scan (in the training set) of the patient with most similar age and gender, and considered the dose volume metrics obtained from automatic plan emulation (Sec.~\ref{sec:generation-ML-dataset}). 

\subsection{Assessment of generalization to clinical plans}\label{sec:method_clinical_validation}
As aforementioned, the 300 plans used to cross-validate our approach were generated with an automatic sampling procedure (Sec.~\ref{sec:artificial-plan-generation}). To assess whether our results on artificial plans can be valid for clinically-used plans, we further observed whether errors obtained for an independent set of ten manually-crafted clinical plans were in line with the errors obtained in the validation of the artificial plans.

To realize the latter step, we trained ML models (as a reminder, one model per OAR - dose-volume metric combination) on the dataset using the 142 CTs and the 300 artificial plans, and reported their prediction errors on a separate set of five CTs associated with two clinical plans each. This was repeated ten times. 

We gathered five clinical plans (three right-sided, two left-sided) for the aforementioned five CTs. Under the supervision of an experienced pediatric radiation oncologist (B.V. Balgobind), two adapted versions of each plan were manually created that both had the isocenter in the middle of the fields. In one plan no block was used and in the other plan a block was introduced to protect part of the liver or spleen, depending on the plan side.

\section{Results}\label{sec:results}

\subsection{Dose-volume metric data distribution}\label{sec:results-distribution}
\begin{figure}
    \centering
    \scalebox{1.0}{
    \setlength{\tabcolsep}{0pt}
    \renewcommand{\arraystretch}{0.0}
    \begin{tabular}{cc}
    
    \includegraphics[width=0.42\linewidth]{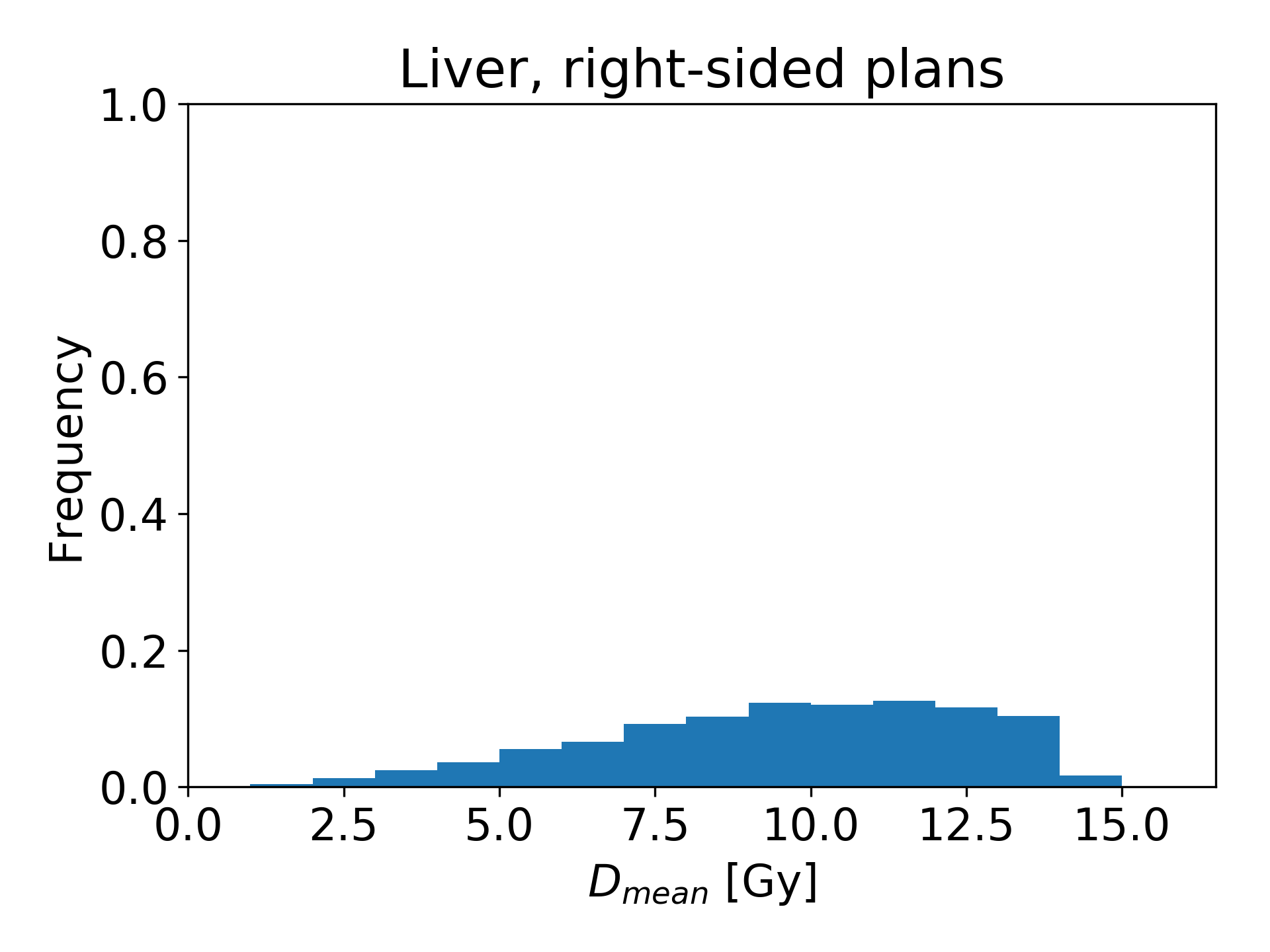} 
    & \includegraphics[width=0.42\linewidth]{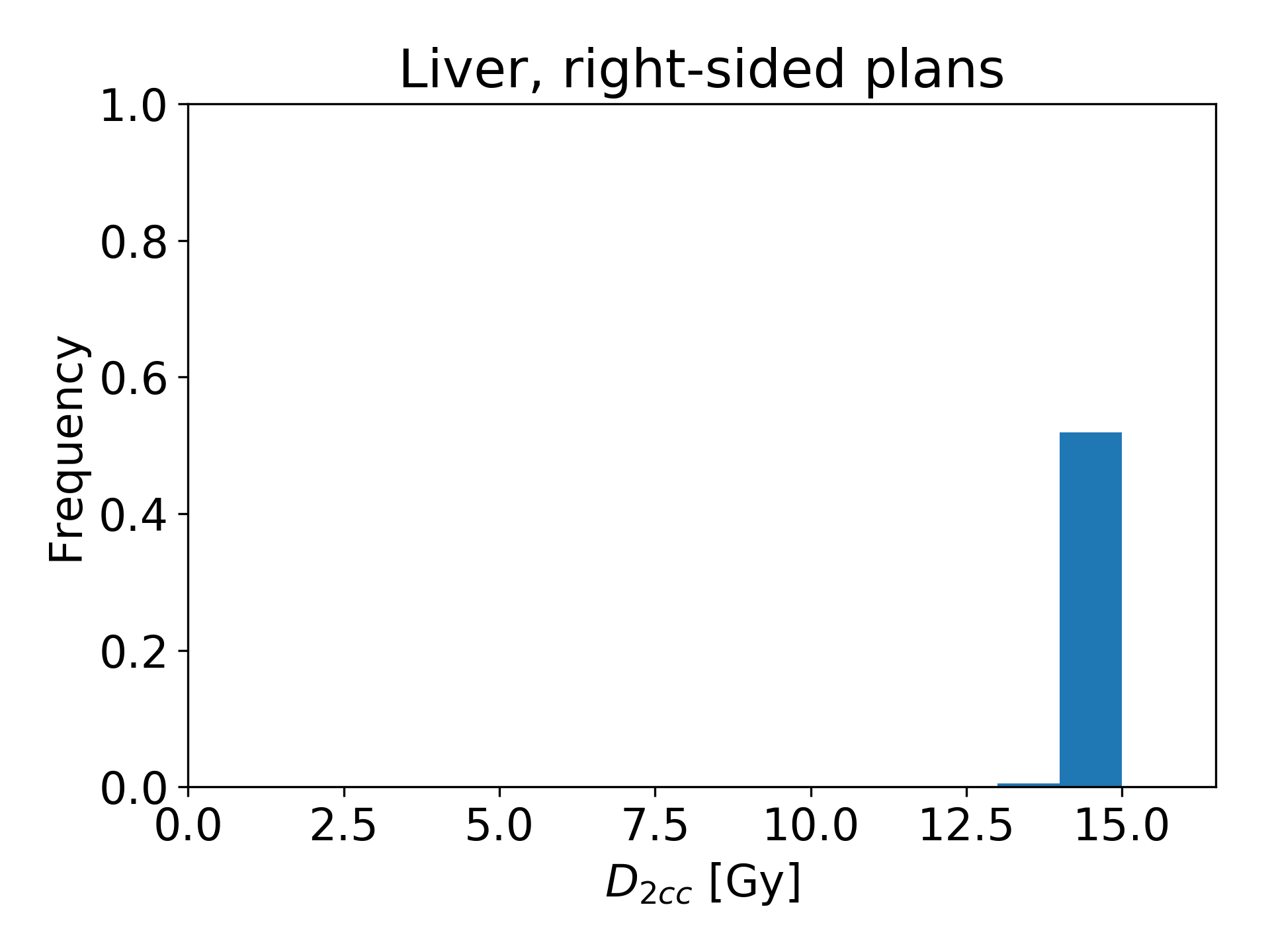} 
    \\
    \includegraphics[width=0.42\linewidth]{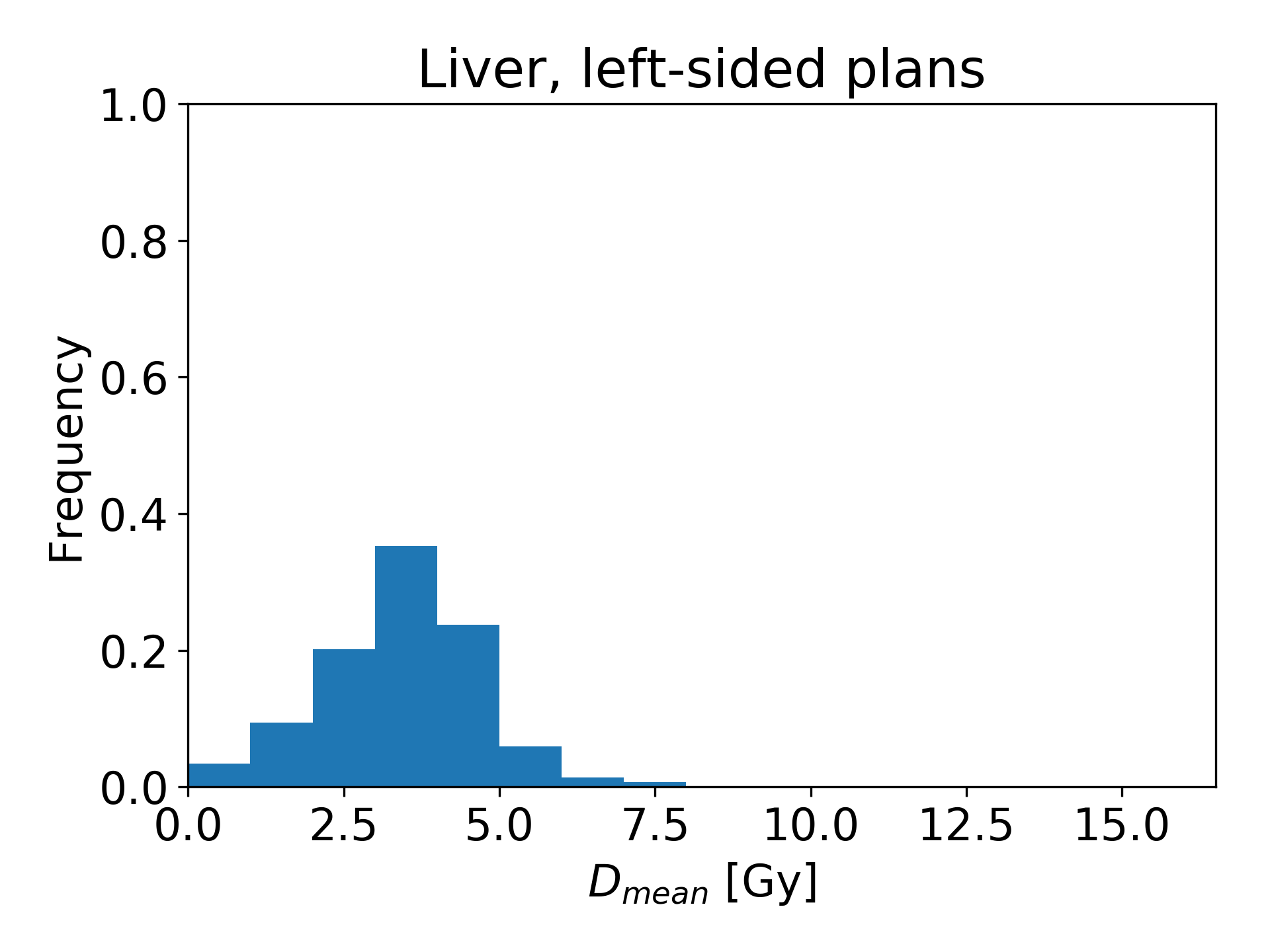} 
    & \includegraphics[width=0.42\linewidth]{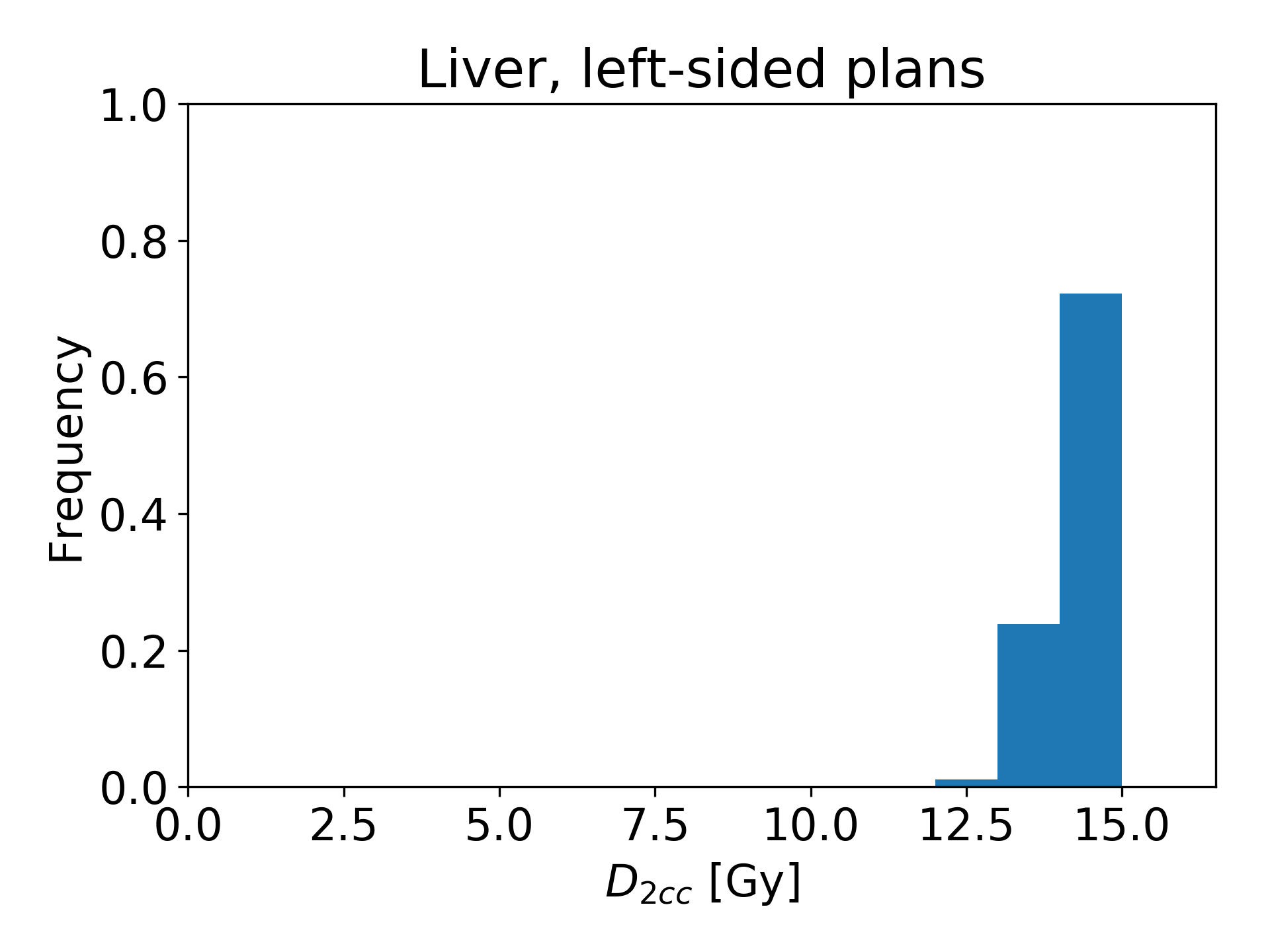} 
    \\
    \includegraphics[width=0.42\linewidth]{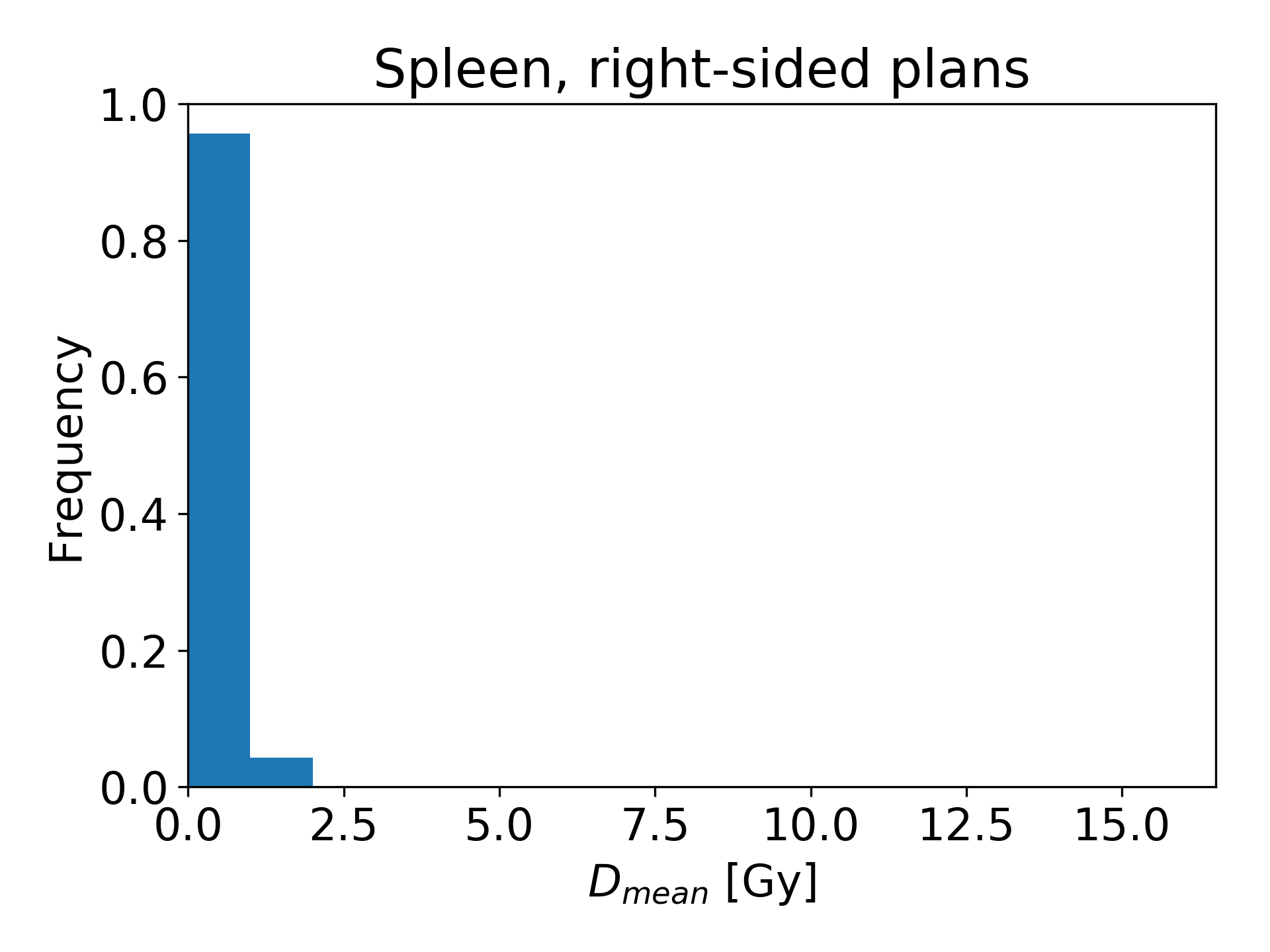} 
    & \includegraphics[width=0.42\linewidth]{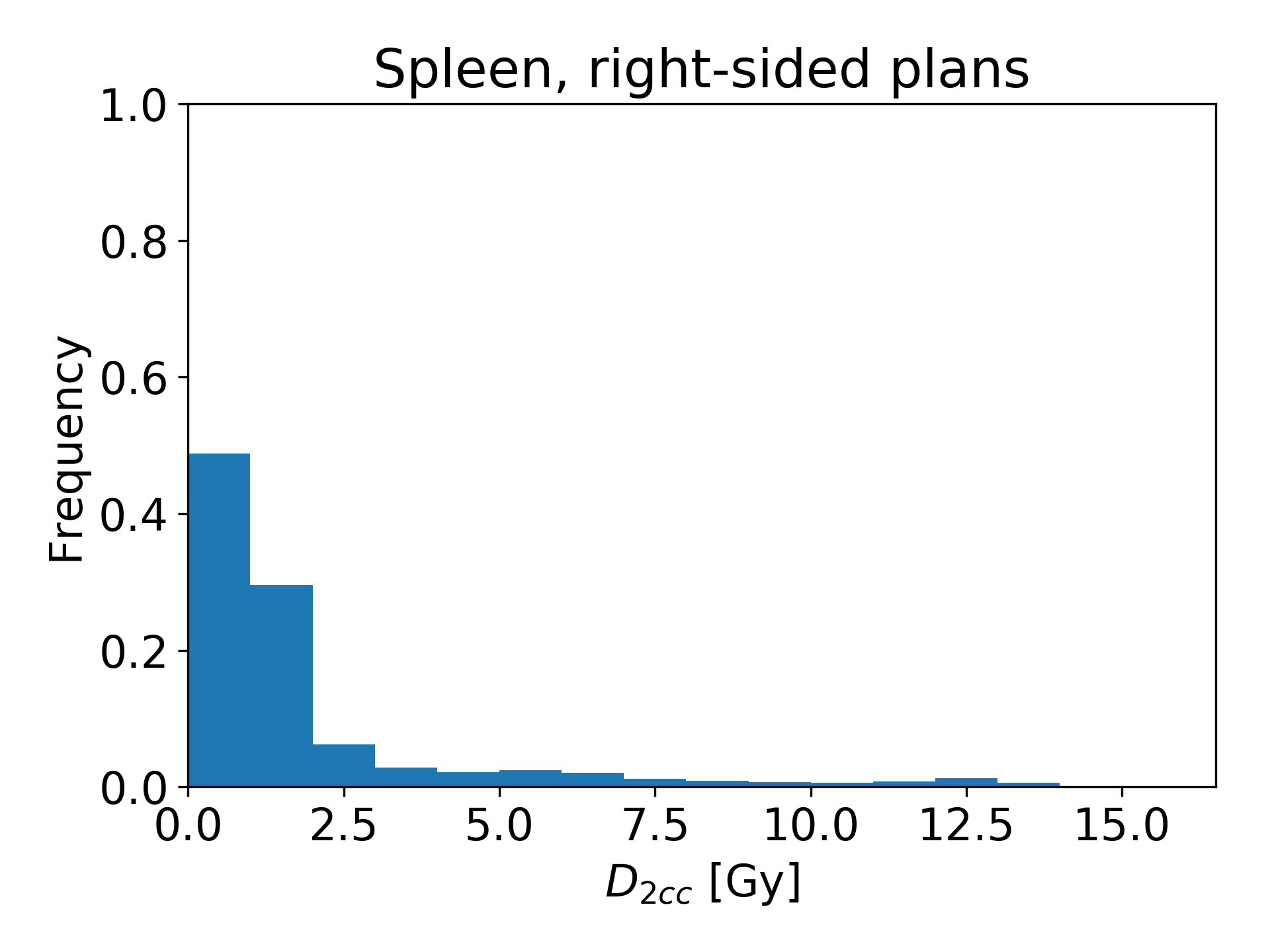} 
    \\
    \includegraphics[width=0.42\linewidth]{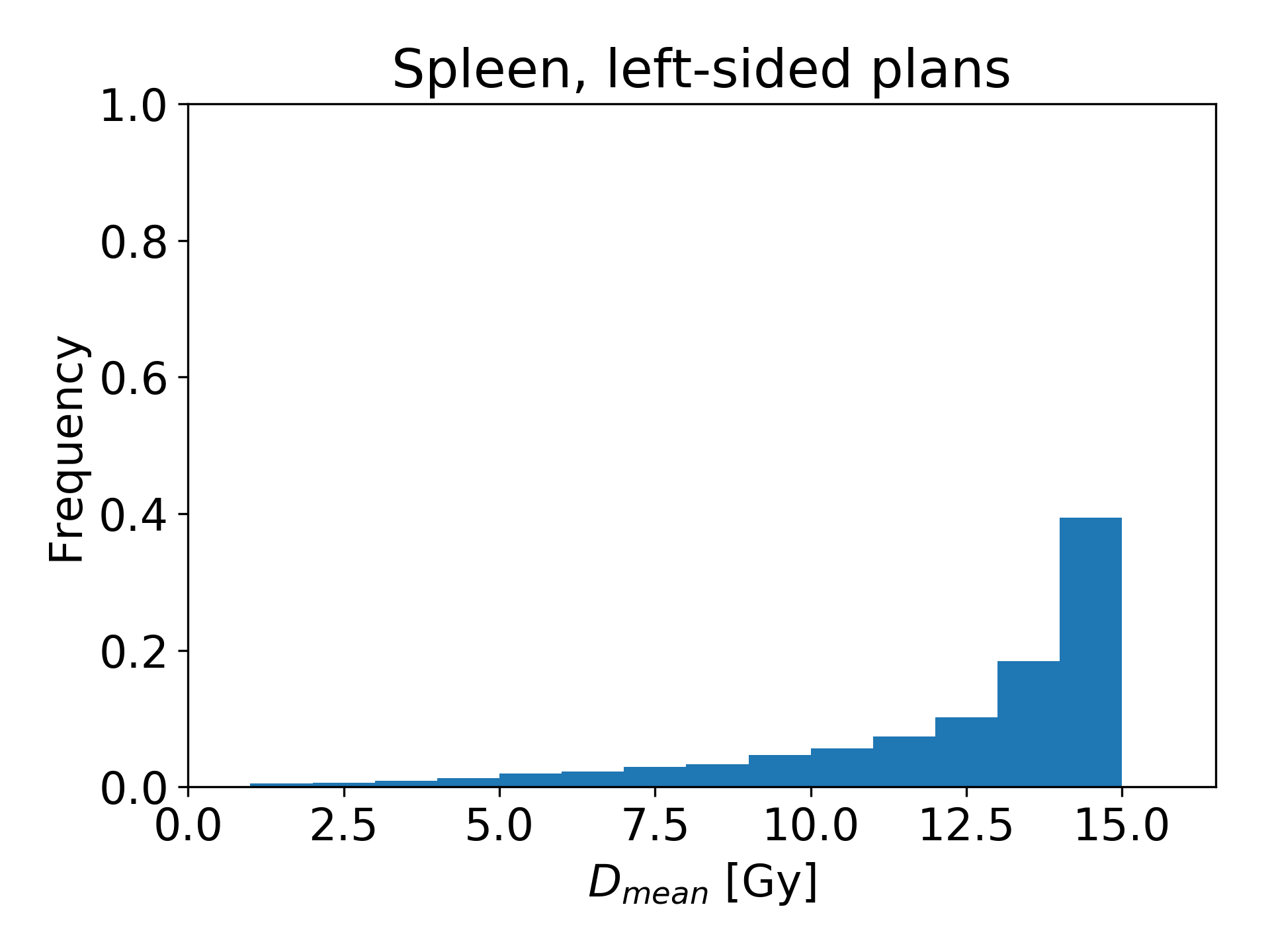} 
    & \includegraphics[width=0.42\linewidth]{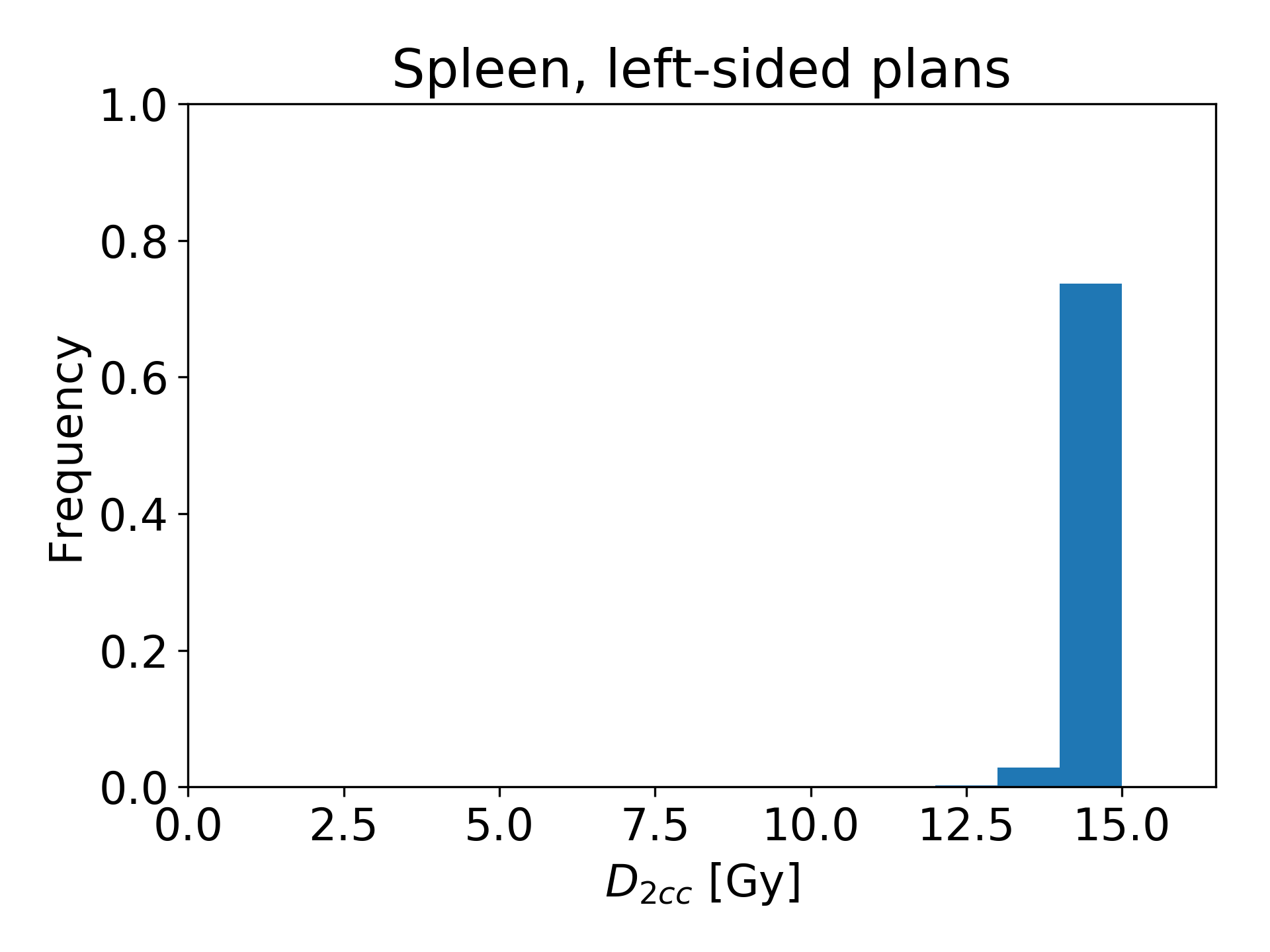} 
    \end{tabular}
    }
    \caption{Distributions for the liver and the spleen of $D\textsubscript{mean}$ and $D\textsubscript{2cc}$ obtained by the automatic plan sampling procedure used to generate artificial plans and by applying the plans to the CT scans.}
    \label{fig:distributions}
\end{figure}

Among the 300 artificial plans, plan side and OAR type was found to influence the distribution of a dose-volume metric considerably.
To illustrate the effect of OAR type and plan side on the dose, Figure~\ref{fig:distributions} shows the distributions found for $D\textsubscript{mean}$ and $D\textsubscript{2cc}$ for the liver and the spleen, in case of left- and right-sided plans. For $D\textsubscript{mean}$ for the liver, distributions approximately resembling the normal distribution were obtained (in case of right-sided plans with particular high variance and long left tail). The distribution in case of right-sided plans had a mean of 9.5 Gy (typically a major part of the liver was in-field), the distribution in case of left-sided plans had a mean of 3.4 Gy (typically a minor part of the liver was in-field). In terms of $D\textsubscript{2cc}$, for the liver we observed values close to the prescribed dose (14.4 Gy) both in case of left- and right-sided plans. The distributions of $D\textsubscript{mean}$ and $D\textsubscript{2cc}$ for the spleen associated with the different plan sides had more marked differences than the ones for the liver. In case of right-sided plans, values close to 0 Gy were obtained for both metrics (typically the spleen was outside the field).
For left-sided plans, large values of $D\textsubscript{mean}$ were found to be much more frequent than low values. The distribution of $D\textsubscript{2cc}$ exhibited a peak around the prescribed dose.
For the contralateral kidney and for the spinal cord the distributions are similar for both plan sides, as the contralateral kidney should be outside the field and the spinal cord should be included within the field (according to protocol).

For all OARs, distributions obtained for $V_\textsubscript{5Gy}$ and $V_\textsubscript{10Gy}$ largely resembled the ones obtained for $D\textsubscript{mean}$. In fact, Pearson correlation coefficients above 98\% were found when comparing $D\textsubscript{mean}$ with $V_\textsubscript{5Gy}$ and $V_\textsubscript{10Gy}$ for almost all OARs. Smaller (yet still large) correlation coefficients were found between $D\textsubscript{mean}$ and $V_\textsubscript{10Gy}$ for the left and right kidney, with values of 96\% and 91\% respectively.

\subsection{Validation on artificial plans}\label{sec:results-artificial}
For each considered OAR, the Mean Absolute Errors (MAEs) (and standard deviation) at validation time for $D\textsubscript{mean}$, $D\textsubscript{2cc}$, $V\textsubscript{5Gy}$, and $V\textsubscript{10Gy}$ from the ten repetitions of the 5-fold cross-validation procedure using the artificial plans are reported in Table~\ref{tab:main-results}. We used statistical significance testing (Wilcoxon rank-sum) to determine whether an outcome is significantly best ($p$-value $< 0.05$).

%%% ONLY ARTIFICIAL NEW
\begin{table}[]
    \caption{Mean test MAE $\pm$ standard deviation of ten repetitions of 5-fold cross-validation for each OAR and dose-volume metric on the artificial plans for the different methods. Bold results are best in that no other method is statistically significantly better.}
    \label{tab:main-results}
    \centering
    \scalebox{1.0}{
    \lineup
    %S[table-format=2.1]S[table-format=2.1]S[table-format=2.1]S[table-format=2.1]

    \begin{tabular}{lll|cccc}
        \br
        \textbf{Side} & \textbf{OAR} & \textbf{Method} & 
         \multicolumn{1}{c}{\textbf{\textit{D}\textsubscript{mean}} [Gy]} &  \multicolumn{1}{c}{\textbf{\textit{D}\textsubscript{2cc}} [Gy]} &  \multicolumn{1}{c}{\textbf{\textit{V}\textsubscript{5Gy}} [\%]} & \multicolumn{1}{c}{\textbf{\textit{V}\textsubscript{10Gy}} [\%]} \\
        \mr
\multirow{12}{*}{ \begin{sideways} \makecell{\textbf{Right}\ \scriptsize (22436 plans)} \end{sideways} } &  & ELN & 1.9$\pm$1.4 &  \bfseries 0.3$\pm$0.2 & 13.3$\pm$10.0 & 13.8$\pm$10.2 \\ 
 & Liver & GPG &  \bfseries 1.7$\pm$1.2 &  \bfseries 0.2$\pm$0.2 &  \bfseries 12.1$\pm$8.7 &  \bfseries 12.7$\pm$9.1 \\ 
 &  & A\&G & 1.9$\pm$1.5 & 0.3$\pm$0.2 & 13.8$\pm$10.9 & 14.3$\pm$11.0 \\ 
\cline{2-7}
 &  & ELN &  \bfseries 0.1$\pm$0.1 &  \bfseries 1.3$\pm$1.7 &  \bfseries 0.7$\pm$0.8 &  \bfseries 0.3$\pm$0.4 \\ 
 & Spleen & GPG & 0.1$\pm$0.1 & 1.6$\pm$2.1 & 0.9$\pm$1.1 & 0.4$\pm$0.6 \\ 
 &  & A\&G & 0.2$\pm$0.2 & 1.7$\pm$2.8 & 1.0$\pm$1.5 & 0.4$\pm$0.8 \\ 
\cline{2-7}
 &  & ELN &  \bfseries 0.5$\pm$0.4 &  \bfseries 2.5$\pm$1.8 &  \bfseries 4.2$\pm$3.7 &  \bfseries 2.5$\pm$2.5 \\ 
 & Kidn.L. & GPG & 0.5$\pm$0.5 & 2.9$\pm$2.0 & 4.4$\pm$4.0 &  \bfseries 2.5$\pm$2.9 \\ 
 &  & A\&G & 0.6$\pm$0.6 & 3.1$\pm$2.8 & 5.2$\pm$5.5 & 3.0$\pm$3.9 \\ 
\cline{2-7} 
 &  & ELN & 0.5$\pm$0.4 &  \bfseries 0.2$\pm$0.2 & 3.8$\pm$3.1 & 3.5$\pm$3.0 \\ 
 & Sp.Cord & GPG &  \bfseries 0.4$\pm$0.4 & 0.2$\pm$0.2 &  \bfseries 3.2$\pm$2.8 &  \bfseries 3.3$\pm$2.8 \\ 
 &  & A\&G & 0.5$\pm$0.5 & 0.3$\pm$0.2 &  \bfseries 3.3$\pm$3.5 & 3.5$\pm$3.6 \\ 
\mr 
\multirow{12}{*}{ \begin{sideways} \makecell{\textbf{Left}\ \scriptsize (20164 plans)} \end{sideways} } &  & ELN & 0.8$\pm$0.7 &  \bfseries 0.3$\pm$0.6 & 6.1$\pm$5.0 & 6.0$\pm$4.8 \\ 
 & Liver & GPG &  \bfseries 0.8$\pm$0.6 &  \bfseries 0.4$\pm$0.6 &  \bfseries 5.8$\pm$4.7 &  \bfseries 5.5$\pm$4.4 \\ 
 &  & A\&G & 1.0$\pm$0.7 & 0.5$\pm$0.9 & 7.7$\pm$5.6 & 7.3$\pm$5.4 \\ 
\cline{2-7} 
 &  & ELN & 1.9$\pm$1.6 &  \bfseries 0.3$\pm$0.7 & 12.9$\pm$11.8 & 14.3$\pm$12.4 \\ 
 & Spleen & GPG & 1.5$\pm$1.3 &  \bfseries 0.3$\pm$0.7 & 9.3$\pm$9.7 & 10.7$\pm$10.2 \\ 
 &  & A\&G &  \bfseries 1.3$\pm$1.5 & 0.4$\pm$0.8 &  \bfseries 8.1$\pm$11.2 &  \bfseries 9.6$\pm$12.0 \\ 
\cline{2-7} 
 &  & ELN & \bfseries 0.2$\pm$0.3 &  \bfseries 1.3$\pm$1.5 & 1.9$\pm$2.4 & 0.8$\pm$1.3 \\ 
 & Kidn.R. & GPG &  0.3$\pm$0.7 &  \bfseries 1.4$\pm$1.6 & 2.4$\pm$4.9 & 0.8$\pm$1.9 \\ 
 &  & A\&G & 0.3$\pm$0.4 &  \bfseries 1.4$\pm$2.1 &  \bfseries 1.8$\pm$3.5 &  \bfseries 0.7$\pm$2.2 \\ 
\cline{2-7} 
 &  & ELN & 0.6$\pm$0.5 &  \bfseries 0.2$\pm$0.2 & 4.9$\pm$4.2 & 4.7$\pm$3.9 \\ 
 & Sp.Cord & GPG &  \bfseries 0.4$\pm$0.3 & 0.2$\pm$0.2 &  \bfseries 3.1$\pm$2.6 &  \bfseries 2.9$\pm$2.3 \\ 
 &  & A\&G & 0.6$\pm$0.4 & 0.3$\pm$0.2 & 3.6$\pm$3.2 & 3.8$\pm$3.1 \\ 
        \br
    \end{tabular}
    }
    \raggedright
    Abbreviations: ELN: Elastic net, GPG: GP-GOMEA, A\&G: Age and gender-based CT scan matching control method, Kidn.L./R.: Kidney Left/Right, Sp.Cord: Spinal cord.
   
    %Bold results are significantly different ($p$-value $< 0.05$) from naive approach of taking mean of training data.
    
\end{table}

Among the three considered methods, the control method based on age and gender-matching (A\&G) performed overall worst, as it was typically found to be statistically significantly inferior to our ML-based approach (either by ELN or GP-GOMEA). The control method often achieved larger MAEs and respective deviations, although dose-volume metric predictions for the spleen and the right kidney for left-sided plans were notable exceptions.
Overall, GP-GOMEA performed better than ELN (among best performing in 18 vs. 14 dose-volume metric - OAR combinations respectively). We therefore proceed by focusing on the results obtained by GP-GOMEA.

The errors for $D\textsubscript{mean}$ and $D\textsubscript{2cc}$ were generally below 2 Gy, which corresponds to approximately 14\% of the prescribed dose of 14.4 Gy. For all OARs but for the spinal cord, the plan side had considerable impact on the magnitude of the errors. As the spinal cord in RL direction was in-field no matter the plan side, the MAEs of dose-volume metrics predictions were found to be small: $< 1$ Gy for $D\textsubscript{mean}$ and $D\textsubscript{2cc}$, $< 4 \%$ for $V\textsubscript{5Gy}$ and $V\textsubscript{10Gy}$. For OARs that were almost out-of-field, e.g., the spleen in case of right-sided plans, small MAEs of $D\textsubscript{mean}$ were found ($< 0.1$ Gy), as very low values were obtained across all patient-plan combinations (see Fig.~\ref{fig:distributions}). 

For the liver in case of right-sided plans, and for the spleen in case of left-sided plans, larger MAEs were found (liver: 1.7 Gy for $D\textsubscript{mean}$, 12.1\% for $V\textsubscript{5Gy}$, 12.6\% for $V\textsubscript{10Gy}$; spleen: 1.5 Gy for $D\textsubscript{mean}$, 9.3\% for $V\textsubscript{5Gy}$, 10.7\% for $V\textsubscript{10Gy}$). These errors can be attributed to the particular configuration of the position of these OARs and the field of the plans.  

Among the dose-volume metrics, $D\textsubscript{2cc}$ for the (partly) in-field OARs had low variability, with a $D\textsubscript{2cc}$ close to the prescribed dose (14.4 Gy). For example, small errors were obtained for the $D\textsubscript{2cc}$ for the liver ($< 0.4$ Gy), as we consistently obtained a large $D\textsubscript{2cc}$ value for both left- and right-sided plans (see Fig.~\ref{fig:distributions}). In contrast, $D\textsubscript{2cc}$ was harder to predict when the OAR was contralateral to the plan side. The MAEs obtained for $D\textsubscript{2cc}$ for the spleen in case of right-sided plans was 1.6 Gy. This was 2.9 Gy for the left kidney, and 1.4 Gy for the right kidney. 

The largest MAE obtained by GP-GOMEA (and also by the other methods) was found for $D\textsubscript{2cc}$ for the left kidney, amounting to 20\% of the prescribed dose. For all dose-volume metrics for the right kidney, and for $D\textsubscript{2cc}$ for the spleen, we found that ML predictions were slightly worse compared to using the baseline (note the negative effect sizes), but not significantly so. Lastly regarding the kidneys, although errors in $D\textsubscript{2cc}$ were relatively large, errors in $V\textsubscript{10Gy}$ were relatively small (compared with $V\textsubscript{10Gy}$ for the other OARs). In fact, only a small percentage of the contralateral kidney, from $0$ to less than $3\%$ typically received at least 10 Gy.

\subsection{Assessment on clinical plans}\label{sec:results-real}
\begin{figure}
    \centering
    \includegraphics[width=\linewidth]{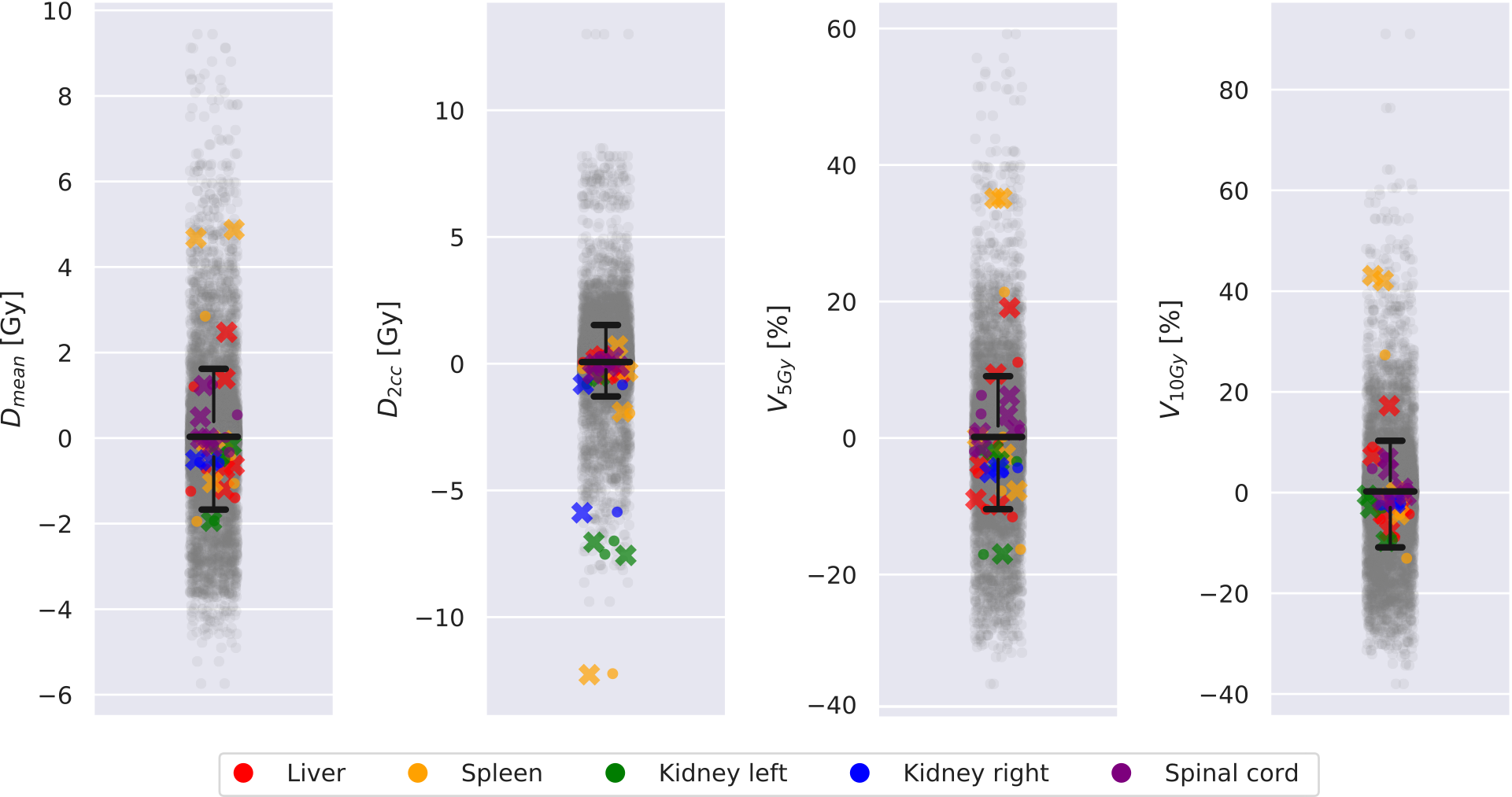}
    \caption{Distribution of signed errors for the ten clinical plans across the dose-volume metrics. Colored dots and crosses represent signed errors for clinical plans without and with blocks respectively. Gray dots represent signed errors from the cross-validation on artificial plans. \hl{Wide} thick black lines represent the median signed error on the artificial plans, while whiskers extend to 1.5 times the interquartile range (default). \hl{Note the use of different scales for the vertical axes.}}
    \label{fig:clinical-distrib}
\end{figure}

Figure~\ref{fig:clinical-distrib} shows the distribution of signed (i.e., non-absolute) errors (as prediction minus actual dose-volume metric) for the independent set of ten clinical cases after training GP-GOMEA (the overall best performing ML algorithm) on the artificial plans. The mean prediction of ten repetitions is considered. To put these errors into perspective, the figure includes the signed errors obtained from the cross-validation on artificial plans (i.e., corresponding to the results of Sec.~\ref{sec:results-artificial}).

Regarding the artificial plans, the figure shows that the errors obtained by GP-GOMEA were unbiased (similar results were found for ELN): no systematic over- nor under-estimations of dose-volume metrics were found on average (note the median error being approximately zero for all metrics). 
Errors obtained for the clinical plans were mostly within the variation that was observed for the artificial plans, with some notable exceptions, e.g., for dose-volume metrics related to the spleen.
The most notable errors are the underestimation of approximately 12 Gy for $D\textsubscript{2cc}$ for both the non-blocked (orange dot) and blocked (orange cross) version of one right-sided clinical plan. Other notable errors are the overestimation of $D\textsubscript{mean}$ and correlated $V\textsubscript{5Gy}$ and $V\textsubscript{10Gy}$ for two left-sided plans with blocks (orange crosses). We provide insights on these aspects in Section~\ref{sec:discussion}. A break down on the errors obtained for the clinical plans is reported in the supplementary material C.

\section{Discussion}\label{sec:discussion}
In this article we presented a new and different paradigm in organ dose reconstruction. By leveraging the modeling power of ML, we showed how patient and plan features can be used to predict organ dose-volume metrics directly, without the need of adopting a surrogate anatomy. Once the ML models are trained, they can readily be used to compute dose-volume metric predictions for a new historical patient and plan, by using their features as input.

Key to obtaining a decent amount of data to perform ML were the collaboration of five international institutes to gather pediatric patient CTs (147), the development of a new automatic sampling procedure yielding artificial Wilms' tumor RT plans, and the creation of an automatic dose reconstruction pipeline to calculate the dose for all patient and plan combinations. We validated our approach on 300 automatically generated artificial plans, and further studied whether the results generalized to ten manually created clinical plans.
Our approach showed promising levels of accuracy in dose reconstruction in both settings. \hl{From the results on the artificial plans, it appeared that the models learned with GP-GOMEA were performing overall slightly better than the ones learned with ELN, yet GP-GOMEA may be harder to implement than ELN. Moreover, while ELN is considered to deliver interpretable models}~\cite{adadi2018peeking,guidotti2018survey}\hl{, for GP-GOMEA this still needs to be validated in clinical practice.}

\begin{figure}
    \centering
    \scalebox{1.0}{
    \setlength{\tabcolsep}{0pt}
    \renewcommand{\arraystretch}{0.0}
    \begin{tabular}{c|c}
    \includegraphics[width=0.44\linewidth,height=0.47\linewidth]{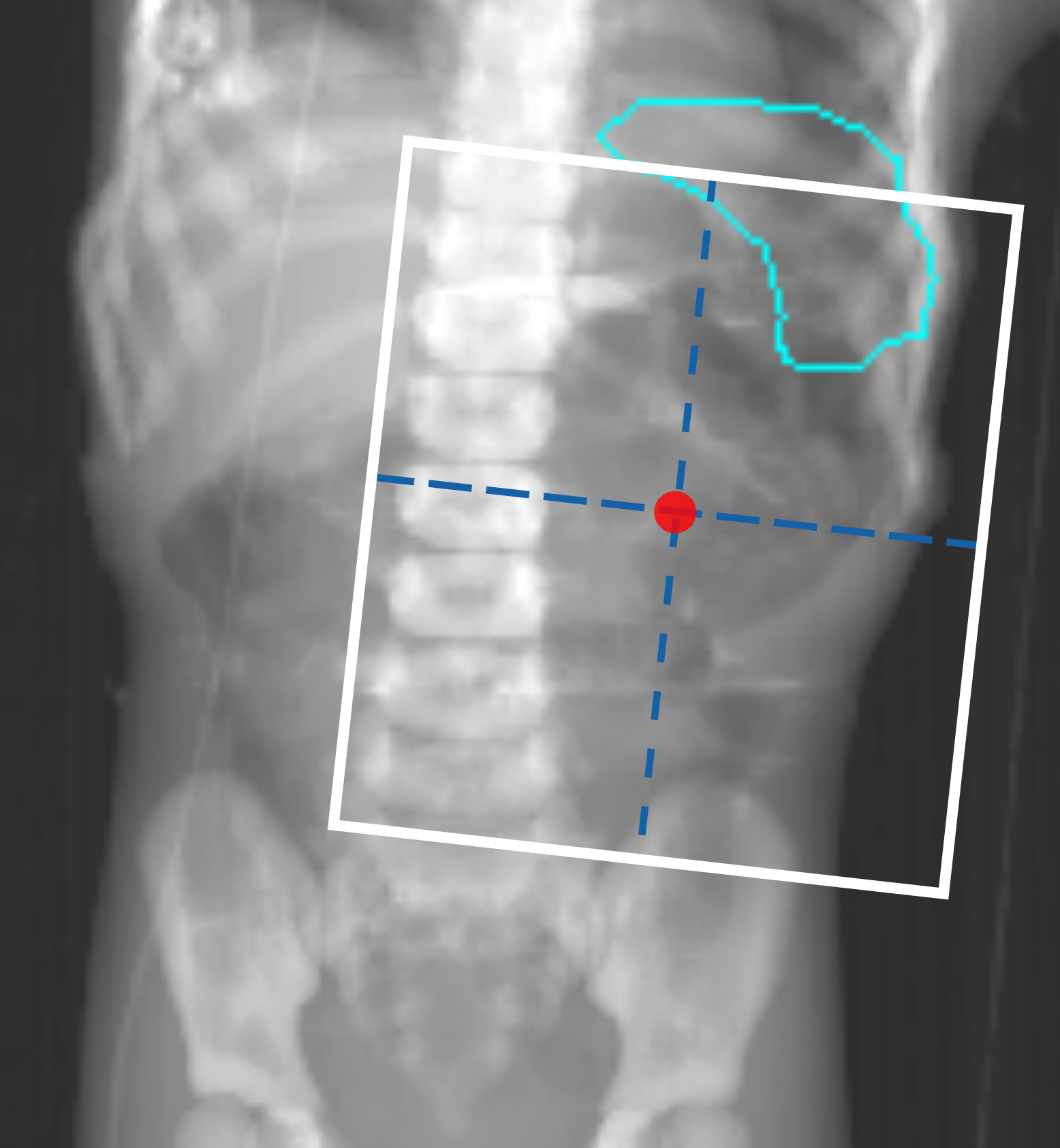}
    & \includegraphics[width=0.44\linewidth,height=0.47\linewidth]{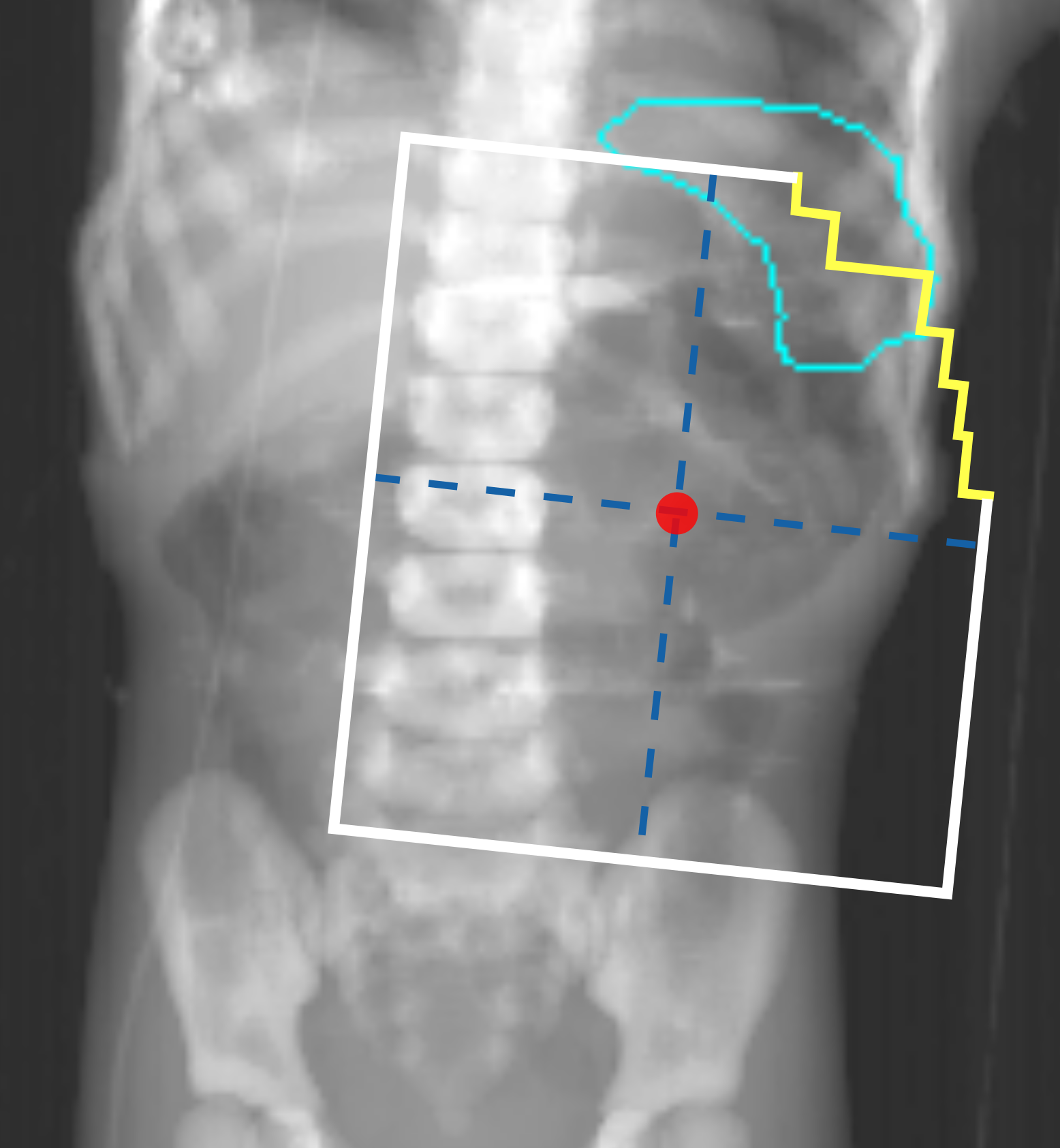} \\
      \includegraphics[width=0.44\linewidth,height=0.47\linewidth]{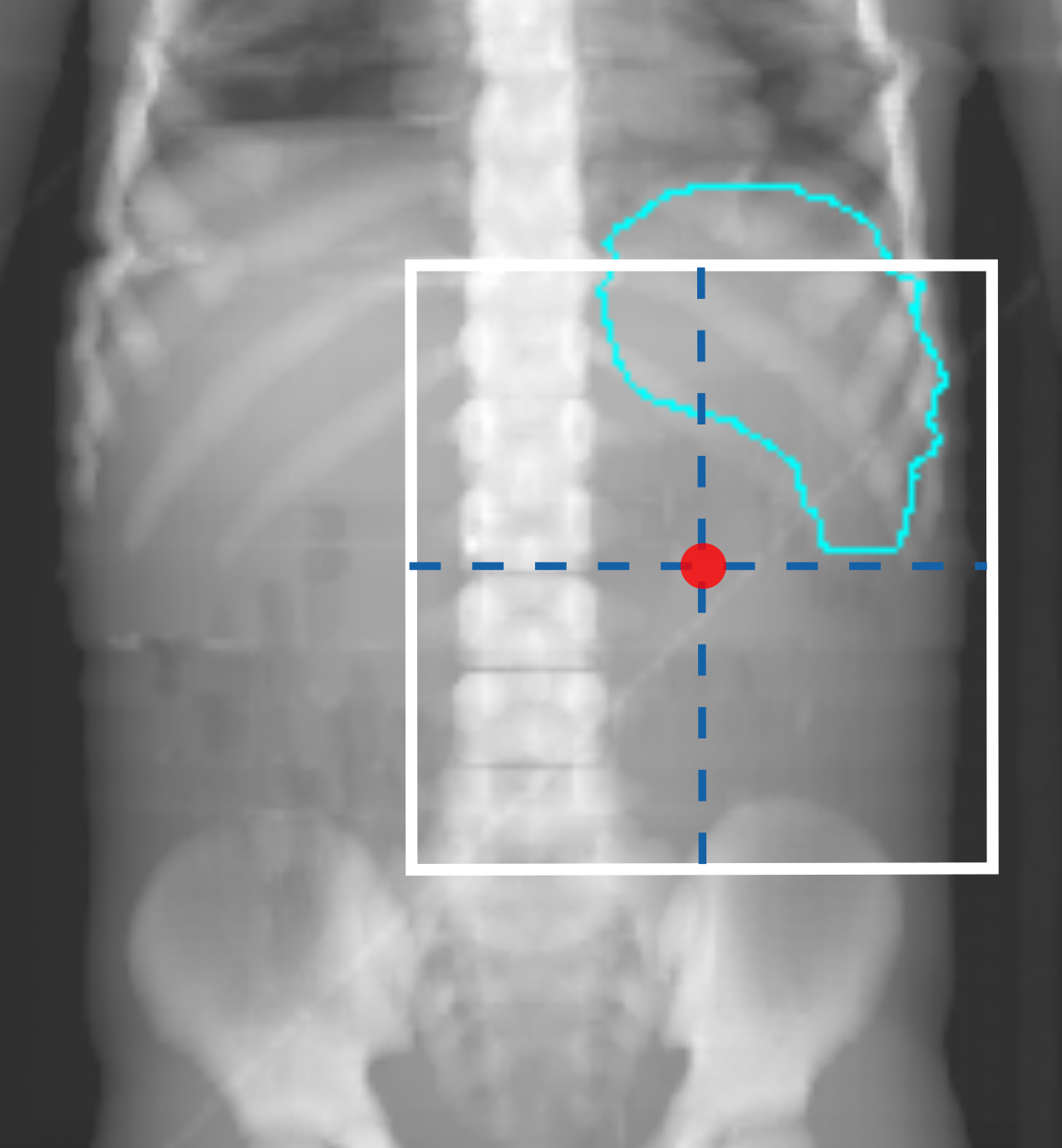}
    & \includegraphics[width=0.44\linewidth,height=0.47\linewidth]{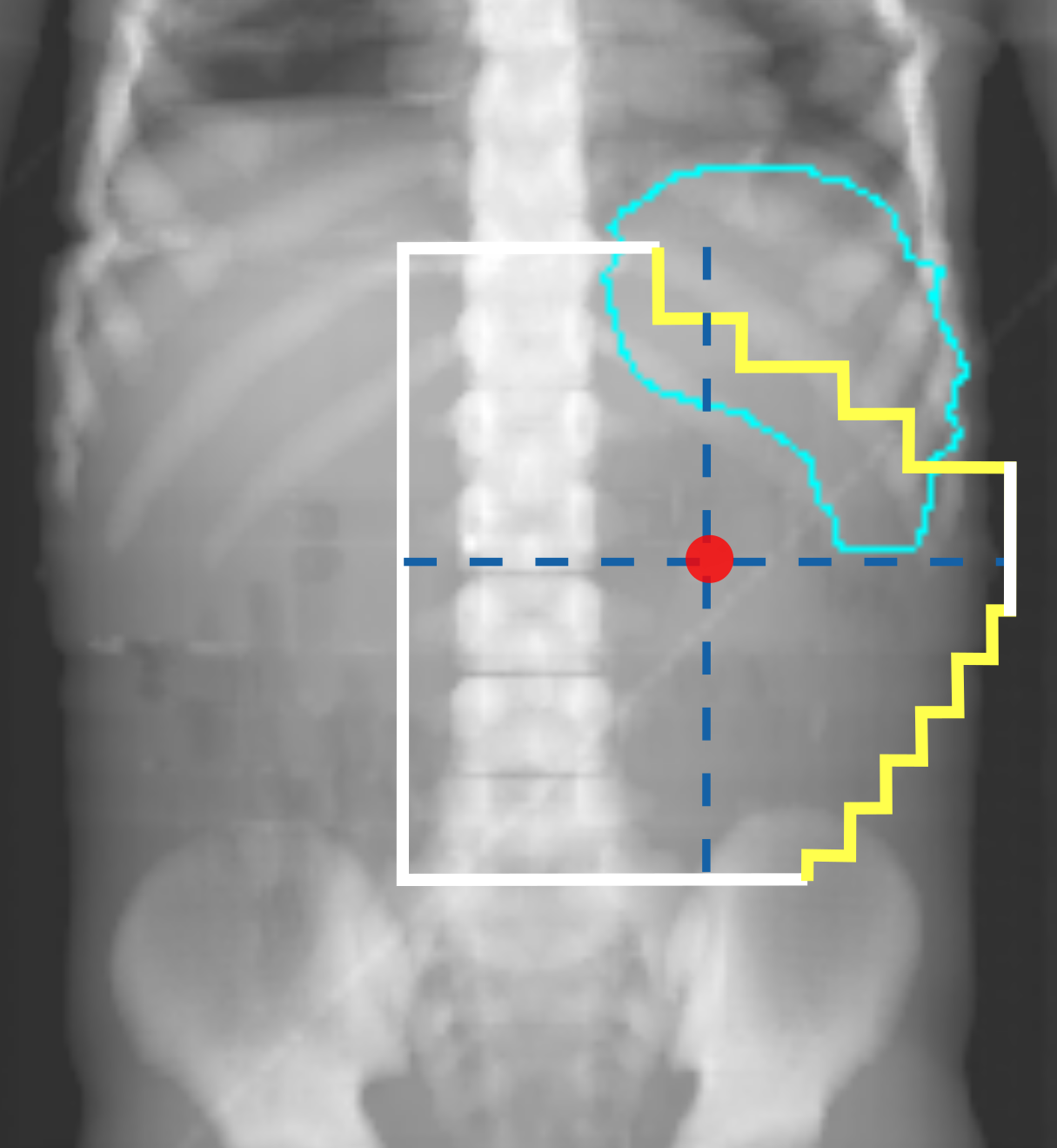} \\
    \end{tabular}
    }
 
    \caption{Effective field shape of two clinical plans without (left column) and with block (right column) plotted on top of the associated DRR. The fields are placed lower than most of the sampled artificial plans (see Fig.~\ref{fig:artificial}). Consequently, for the plans with blocks, our approach produced an overestimation of 5 Gy for the $D\textsubscript{mean}$ of the spleen (see the two orange crosses in Fig.~\ref{fig:clinical-distrib}). }
    \label{fig:outlier_plan}
\end{figure}

For some metrics and OARs, errors were relatively large for some of the clinical plans.
This may be due to chance, because ten is a small number.
For example, the large underestimation observed for the $D\textsubscript{2cc}$ of the spleen for one clinical case (with and without block) is due to the ML-based approach wrongly predicting a value associated with a spleen that is located completely outside of the field.
Another reason why errors were relatively large for the clinical plans is that the artificial plan generation method needs to be improved. Artificial plans were generated by sampling geometry properties \emph{uniformly} within predefined boundaries on two reference DRRs. Uniform sampling might not be representative of the distribution clinical plans have. Moreover, we consulted a single radiation oncologist to define clinically acceptable boundaries to use in the sampling of artificial plans. Consulting multiple experts and allowing for a larger variation might better help covering the extent of variation that is present in historical plans (Sec.~\ref{sec:artificial-plan-generation}). For example, the isocenter locations of artificial left-sided plans were never sampled below the 1st lumbar vertebra (see Fig.~\ref{fig:artificial}) and approximately half of the $D\textsubscript{mean}$ values for the spleen in case of the artificial left-sided plans were close to the prescribed dose (14.4 Gy, see Fig.~\ref{fig:distributions}), which means that the spleen was often almost completely in-field in our artificially generated set of left-sided plans. When a block was applied, only a small part of the spleen was spared. However, in clinical practice, isocenter locations can be lower, and a larger part of the spleen might actually be outside the field (see Fig.~\ref{fig:outlier_plan}). This might explain the relatively large errors observed in the two outlier cases in Figure~\ref{fig:clinical-distrib} where the isocenter location of the clinical plans is lower than the sampled range. Ultimately, effort should be done to improve the sampling of artificial plans.

In the validation performed upon artificial plans as well as in the assessment concerning clinical plans, a main result that emerges is that dose-volume metrics for an organ are hard to predict when, due to the field setup, it is unclear whether the OAR is (partially) included in the field or not. For example, the $D\textsubscript{2cc}$ is very different when a tiny part of an OAR is inside the field compared to when the OAR is completely outside the field.
As experimentally observed in prior work \cite{virgolin2018feasibility,virgolin2019machine}, 2D bony anatomy provides only coarse information on OAR shape and position even for ML algorithms (e.g., an MAE of 6.4 mm for the prediction of the liver position along the IS axis was reported \cite{virgolin2019machine}). Yet, because bony anatomy is the only structure that is reliably visible in historical radiographs, most of the anatomical features rely on it. Patients with similar anatomical features derived from bony anatomy may have different OAR shape and position, and thus different dose-volume metrics. Furthermore, impreciseness in feature values due to e.g., uncertainties in landmark detection and plan emulation, aggravate the situation.

Compared to conventional dose reconstruction methods (that use surrogate anatomies and heuristics to decide what surrogate to use), we considered a relatively large number of features: 33. Phantom-based methods consider, e.g., only age and gender \cite{stovall2006dose,howell2019adaptations}, or gender together with height and weight percentiles \cite{geyer2014uf}. However, if a 2D radiograph is available, the added value of this information should be exploited. In our work, the majority of the features we considered, i.e., 23 out of 33 (minus eight due to automatic feature selection, see supplementary A), regarded patient anatomy as visible on a 2D radiograph, which we simulated with DRRs. Our DRRs were generated in a conformal fashion, e.g., the abdomen was always fully included in RL direction. The automatic landmark detection that was used to generate features expects this conformity to achieve precise detection \cite{wang2020spiej}. When dealing with actual historical radiographs, however, several challenges need to be taken into account. For example, our automatic landmark detection method requires further development to account for noise in the radiograph (e.g., the presence of hand-writing on the radiograph). Moreover, educated guesses of landmark locations may be needed in some cases, as some historical radiographs do not include the entire abdomen (see Fig.~\ref{fig:historicalplans}(c)). Nevertheless, as long as the features are somehow collected (e.g., manually), they can be used as input for the ML models to get respective dose-volume metric predictions.

There are disadvantages of our approach compared to conventional dose reconstruction methods that use surrogate anatomies beyond the need for patient radiographs, which are not always available in retrospective data. In particular, a key limitation is that ML models do not predict the entire 3D dose distribution an organ receives, but only the metrics they were trained for. Potentially useful information to link to AEs may be contained in 3D dose distributions. To predict 3D dose distributions, the ML models would need to be trained to predict a 3D output. Surrogate-based methods do allow to obtain the entire 3D dose distribution to an organ, since the distribution can be visualized on the organ of the surrogate anatomy, after plan simulation. However, considering the magnitude and variations of the errors of organ mean dose obtained by conventional approaches \cite{wang2018age}, it is questionable whether the full 3D distribution will be sufficiently reliable. Our approach, as currently proposed, can straightforwardly be extended to predict any (scalar) dose-volume metric that is suspected to be useful to study AEs (it suffices to train ML on that metric).

Another limitation of our approach is that it does not take into consideration uncertainties related to OAR motion. For validation, we aimed at reconstructing the dose based on the particular snapshot of anatomy at the moment the CT (ground-truth) / respective DRR (to simulate historical radiographs) was taken. Yet, OAR motion plays a key role in the uncertainty of organ positioning at the edge of the field, which can lead to a discrepancy between the planned dose and the actual delivered dose.
In RT practice, radiation delivery is performed over a number of days, with fractionation schemes. The OAR position can therefore vary (i.e., inter-fractional position variation). Intra-fractional organ motion due to, e.g., respiration variation, contributes to the difference between planned dose and delivered dose as well \cite{huijskens2015quantification}.

Lastly, a main limitation of our approach is that the ML models we generated are specific to pediatric patients (1 to 8 years of age) and Wilms' tumor RT plans: they can only predict reliable dose-volume metrics of specific OARs they were explicitly trained for. The RT plans we have sampled were also restricted to a standard AP-PA setup without considering wedges, boost fields, or other radiation sources such as Cobalt-60. Moreover, the predictions of the ML models (as well as the validation performed in this study) are based on the dose calculation algorithm we adopted when preparing training data, which has inaccuracies. Specifically, we used a collapsed cone algorithm available in Oncentra TPS. Though good accuracy was reported in the in-field and near-field region ($<$ 5 cm from the field borders, achieves an error of 1-2\% of the prescribed dose), in low dose regions (10--15 cm from the field border) an underestimation of 10\% of the dose in the region was reported \cite{krieger2005monte}. We remark that the OARs we considered in this study were mostly within 5 cm near the field border (except for the spleen in case of the right-sided plans). To make the method more general for OARs far from field borders, more advanced Monte Carlo dose calculation algorithms should be applied in future implementations. 

We remark that our approach could be readily extended to predict dose-volume metrics for other types of RT plans that use abdominal fields by making use of our feature extraction method based on landmark detection (Sec.~\ref{sec:features-used}), which is general to the abdominal area. Furthermore, we believe that the core ideas of our work can be replicated for other cohorts and other types of plans. Essentially, as long as a sufficient number of anatomies and plans are collected or generated, and a large number of dose reconstructions are performed to be used as examples, new ML models can be trained to predict how the dose-volume metrics are linked to anatomy-plan configurations. Beyond this, our algorithm GP-GOMEA is a general machine learning method that can be used for arbitrary tasks that require a form of regression. As was the case in our study, the collection and preparation of sufficient data for ML is likely to be the largest required effort.

Our proposed approach presents several advantages compared to traditional dose reconstruction methods. First of all, the validation results on artificial plans showed that our approach is superior to the control method based on age and gender-matching, which simulates the conventional way of surrogate-based dose reconstruction methods. 
Note that the control method can be considered to operate in an advantaged setting because the plan of interest is emulated on the surrogate CT (whereas ML needs to infer salient information by solely relying on plan features).

We also found our validation results to compare favorably with respect to our recent work considering dose reconstruction for a similar childhood cancer cohort \cite{wang2018age}. Table~\ref{tab:literature-results} 
summarizes several statistics of organ mean dose errors for liver and spleen  based on the validation results of our approach and the values reported in~\cite{wang2018age}. It can be observed that our approach achieved both smaller prediction errors and smaller interquartile ranges for signed errors. We remark, however, that since the dose reconstruction accuracy is largely influenced by the particular plans considered, these values may not be a fair comparison.

\begin{table}[]
    \caption{A summary of validation results for the $D_\textsubscript{mean}$ of liver and spleen concerning~\cite{wang2018age} and this study.}
    \label{tab:literature-results}
    \centering
    \small
    \lineup
    \begin{tabular}{ccccccc}
    \mr
\textbf{Study} & \tabincell{c}{\textbf{No. CTs} \\ \textbf{included}} &  \tabincell{c}{\textbf{Patient} \\\textbf{age} (yrs)} 
& \tabincell{c}{\textbf{No. plans} \\ \textbf{included} }  & \textbf{OAR} & \tabincell{c}{\textbf{MAE} \\ (Gy)}  & \tabincell{c}{\textbf{IQR} \\ (Gy) }  \\ 
\mr
 \multirow{2}{*}{ Wang et al 2018 } &  \multirow{2}{*}{31} &  \multirow{2}{*}{2--5} & \multirow{2}{*}{12} & Liver &  12 & 3.6 \\

 &  & & &  Spleen & 2.6 & 4.7 \\ 

 \multirow{2}{*}{ This, GPG\textsuperscript{A} } & \multirow{2}{*}{142} & \multirow{2}{*}{1--8} & \multirow{2}{*}{300} & Liver & 1.3 & 2.0  \\ 
  & & & & Spleen & 0.8 & 1.2\\ 
 \multirow{2}{*}{ This, GPG\textsuperscript{C} } & \multirow{2}{*}{5} & \multirow{2}{*}{1--8} & \multirow{2}{*}{10} & Liver & 1.1 & 1.9  \\ 
 & & & & Spleen & 1.7 & 2.2\\ 
 \br
    \end{tabular}

    \raggedright
    Abbreviations: IQR: interquantile range, GPG\textsuperscript{A}: GP-GOMEA on artificial plans, GPG\textsuperscript{C}: GP-GOMEA on clinical plans.
\end{table}

We are currently working on a multi-institute study to compare our approach with two state-of-the-art, phantom-based dose reconstruction approaches \cite{lee2015reconstruction,howell2019adaptations}. In that study, a same set of patients and plans will be used for validation.  

Finally, a benefit of having ML models is that, once features are collected, they can be used as inputs for the model to obtain the prediction of a dose-volume metric immediately. Running a model on a computer simply means to follow the steps encoded by the formula the model represents, which takes a few milliseconds. Conversely, in a surrogate-based approach (e.g., in the age and gender-matching), the features are used to craft or select a surrogate anatomy. Then, effort and time must be put to emulate the plan on the surrogate anatomy, calculate the dose, and obtain the dose-volume metrics \cite{lee2015reconstruction,howell2019adaptations,wang2020spiej}.

\section{Conclusion}
We presented the first surrogate-free organ dose reconstruction method based on ML.
Our method was enabled by the collection of large amounts of patient and CT data, and the automatic generation of artificial plans and of dose distribution data. We assembled a dataset of dose-volume metrics corresponding to features of patient anatomy and plan geometry, and subsequently trained ML models to predict how features of patient anatomy and of treatment plans influence dose-volume metrics. The predictions were validated upon both artificial and clinical RT plans, and achieved good accuracy in both cases.

\section*{Acknowledgements}
The authors acknowledge Stichting Kinderen Kankervrij (KiKa) for financial support (project \#187), and the Maurits and Anna de Kock foundation for financing a high performance computing system. Elekta is acknowledged for providing the research software ADMIRE for automatic segmentation. The authors thank Abigail Bryce-Atkinson for her help in data preparation and feature extraction, and Dr. C\' ecile M. Ronckers for sharing her expertise.

\section*{References}

\bibliographystyle{custom_jphysicsB}
\bibliography{references}

@article{cheung2017chronic,
  title={Chronic health conditions and neurocognitive function in aging survivors of childhood cancer: A report from the Childhood Cancer Survivor Study},
  author={Cheung, Yin Ting and Brinkman, Tara M and Li, Chenghong and Mzayek, Yasmin and Srivastava, Deokumar and Ness, Kirsten K and Patel, Sunita K and Howell, Rebecca M and Oeffinger, Kevin C and Robison, Leslie L and others},
  journal={J. Nat. Cancer. Inst.},
  volume={110},
  number={4},
  pages={411--419},
  year={2017},
  publisher={Oxford University Press}
}

@article{hoerl1970ridge,
  title={Ridge regression: {B}iased estimation for nonorthogonal problems},
  author={Hoerl, Arthur E and Kennard, Robert W},
  journal={Technometrics},
  volume={12},
  number={1},
  pages={55--67},
  year={1970},
  publisher={Taylor \& Francis Group}
}

@article{tibshirani1996regression,
  title={Regression shrinkage and selection via the lasso},
  author={Tibshirani, Robert},
  journal={J. R. Stat. Soc. B},
  volume={58},
  number={1},
  pages={267--288},
  year={1996},
  publisher={Wiley Online Library}
}

@article{zou2005regularization,
  title={Regularization and variable selection via the elastic net},
  author={Zou, Hui and Hastie, Trevor},
  journal={J. R. Stat. Soc. B},
  volume={67},
  number={2},
  pages={301--320},
  year={2005},
  publisher={Wiley Online Library}
}

@article{howell2019adaptations,
    author = {Rebecca M. Howell and Susan A. Smith and Rita E. Weathers and Stephen F. Kry and Marilyn Stovall},
    title = {Adaptations to a Generalized Radiation Dose Reconstruction Methodology for Use in Epidemiologic Studies: An Update from the {MD} {A}nderson Late Effect Group},
    volume = {192},
    journal = {Radiat. Res.},
    number = {2},
    publisher = {Radiation Research Society},
    pages = {169--188},
    year = {2019},
}

@article{birgisson2005adverse,
  title={Adverse effects of preoperative radiation therapy for rectal cancer: long-term follow-up of the {S}wedish {R}ectal {C}ancer {T}rial},
  author={Birgisson, Helgi and P\r{a}hlman, Lars and Gunnarsson, Ulf and Glimelius, Bengt},
  journal={J. Clin. Oncol.},
  volume={23},
  number={34},
  pages={8697--8705},
  year={2005},
  publisher={American Society of Clinical Oncology}
}

@article{van2010evaluation,
  title={Evaluation of late adverse events in long-term {W}ilms' tumor survivors},
  author={van Dijk, Irma W. E. M. and Oldenburger, Foppe and Cardous-Ubbink, Mathilde C and Geenen, Maud M and Heinen, Richard C and de Kraker, Jan and van Leeuwen, Flora E and van der Pal, Helena JH and Caron, Huib N and Koning, Caro CE and others},
  journal={Int. J. Radiat. Oncol. Biol. Phys.},
  volume={78},
  number={2},
  pages={370--378},
  year={2010},
  publisher={Elsevier}
}

@article{feng2007intensity,
  title={Intensity-modulated radiotherapy of head and neck cancer aiming to reduce dysphagia: early dose--effect relationships for the swallowing structures},
  author={Feng, Felix Y and Kim, Hyungjin M and Lyden, Teresa H and Haxer, Marc J and Feng, Mary and Worden, Frank P and Chepeha, Douglas B and Eisbruch, Avraham},
  journal={Int. J. Radiat. Oncol. Biol. Phys.},
  volume={68},
  number={5},
  pages={1289--1298},
  year={2007},
  publisher={Elsevier}
}

@article{donovan2007randomised,
  title={Randomised trial of standard 2{D} radiotherapy ({RT}) versus intensity modulated radiotherapy ({IMRT}) in patients prescribed breast radiotherapy},
  author={Donovan, Ellen and Bleakley, Natalie and Denholm, Erica and Evans, Phil and Gothard, Lone and Hanson, Jane and Peckitt, Clare and Reise, Stephanie and Ross, Gill and Sharp, Grace and others},
  journal={Radiother. Oncol.},
  volume={82},
  number={3},
  pages={254--264},
  year={2007},
  publisher={Elsevier}
}

@article{bolling2011dose,
  title={Dose--Volume Analysis of Radiation Nephropathy in Children: Preliminary Report of the Risk Consortium},
  author={B{\"o}lling, Tobias and Ernst, Iris and Pape, Hildegard and Martini, Carmen and R{\"u}be, Christian and Timmermann, Beate and Fischedick, Karin and Kortmann, Rolf-Dieter and Willich, Normann},
  journal={Int. J. Radiat. Oncol. Biol. Phys.},
  volume={80},
  number={3},
  pages={840--844},
  year={2011},
  publisher={Elsevier}
}

@article{leisenring2009pediatric,
  title={Pediatric cancer survivorship research: experience of the Childhood Cancer Survivor Study},
  author={Leisenring, Wendy M and Mertens, Ann C and Armstrong, Gregory T and Stovall, Marilyn A and Neglia, Joseph P and Lanctot, Jennifer Q and Boice Jr, John D and Whitton, John A and Yasui, Yutaka},
  journal={J. Clin. Oncol.},
  volume={27},
  number={14},
  pages={2319},
  year={2009},
  publisher={American Society of Clinical Oncology}
}

@article{emami2013tolerance,
author={Emami, B.
and Lyman, J.
and Brown, A.
and Cola, L.
and Goitein, M.
and Munzenrider, J. E.
and Shank, B.
and Solin, L. J.
and Wesson, M.},
title={Tolerance of normal tissue to therapeutic irradiation},
journal={Int. J. Radiat. Oncol. Biol. Phys.},
year={1991},
month={May},
day={15},
publisher={Elsevier},
volume={21},
number={1},
pages={109-122}
}

@article{constine2019pediatric,
  title={{P}ediatric {N}ormal {T}issue {E}ffects in the {C}linic ({PENTEC}): {A}n International Collaboration to Analyse Normal Tissue Radiation Dose--Volume Response Relationships for Paediatric Cancer Patients},
  author={Constine, LS and Ronckers, C M and Hua, C-H and Olch, A and Kremer, L C M and Jackson, A and Bentzen, S M},
  journal={Clin. Oncol.},
  volume={31},
  number={3},
  pages={199--207},
  year={2019},
  publisher={Elsevier}
}

@article{verellen2008short,
  title={A (short) history of image-guided radiotherapy},
  author={Verellen, Dirk and De Ridder, Mark and Storme, Guy},
  journal={Radiother. Oncol.},
  volume={86},
  number={1},
  pages={4--13},
  year={2008},
  publisher={Elsevier}
}

@article{stovall2006dose,
  title={Dose reconstruction for therapeutic and diagnostic radiation exposures: use in epidemiological studies},
  author={Stovall, Marilyn and Weathers, Rita and Kasper, Catherine and Smith, Susan A and Travis, Lois and Ron, Elaine and Kleinerman, Ruth},
  journal={Radiat. Res.},
  volume={166},
  number={1},
  pages={141--157},
  year={2006}
}

@article{ng2012individualized,
  title={Individualized {3D} reconstruction of normal tissue dose for patients with long-term follow-up: a step toward understanding dose risk for late toxicity},
  author={Ng, Angela and Brock, Kristy K and Sharpe, Michael B and Moseley, Joanne L and Craig, Tim and Hodgson, David C},
  journal={Int. J. Radiat. Oncol. Biol. Phys.},
  volume={84},
  number={4},
  pages={e557--e563},
  year={2012},
  publisher={Elsevier}
}

@article{lee2015reconstruction,
  year = 2015,
    month = {feb},
    publisher = {{IOP} Publishing},
    volume = {60},
    number = {6},
    pages = {2309--2324},
    author = {Choonik Lee and Jae Won Jung and Christopher Pelletier and Anil Pyakuryal and Stephanie Lamart and Jong Oh Kim and Choonsik Lee},
    title = {Reconstruction of organ dose for external radiotherapy patients in retrospective epidemiologic studies},
    journal = {Phys. Med. Biol.}
}

@article{bezin2017review,
  title={A review of uncertainties in radiotherapy dose reconstruction and their impacts on dose--response relationships},
  author={Bezin, J{\'e}r{\'e}mi V{\~u} and Allodji, Rodrigue S and M{\`e}ge, Jean-Pierre and Beldjoudi, Guillaume and Saunier, Fleur and Chavaudra, Jean and Deutsch, Eric and de Vathaire, Florent and Bernier, Val{\'e}rie and Carrie, Christian and others},
  journal={J. Radiol. Prot.},
  volume={37},
  number={1},
  pages={R1},
  year={2017},
  publisher={IOP Publishing}
}

@article{xu2014exponential,
	year = 2014,
	month = {aug},
	publisher = {{IOP} Publishing},
	volume = {59},
	number = {18},
	pages = {R233--R302},
	author = {X George Xu},
	title = {An exponential growth of computational phantom research in radiation protection, imaging, and radiotherapy: a review of the fifty-year history},
	journal = {Phys. Med. Biol.}
}

@article{icrp89,
  title={Basic anatomical and physiological data for use in radiological protection: reference values: {ICRP} Publication 89},
  author={Valentin, Jack},
  journal={Annals of the ICRP},
  volume={32},
  number={3-4},
  pages={1--277},
  year={2002},
  publisher={Elsevier}
}

@article{wang2018age,
  title={Are age and gender suitable matching criteria in organ dose reconstruction using surrogate childhood cancer patients' {CT} scans?},
  author={Wang, Z. and van Dijk, Irma W. E. M. and Wiersma, J. and Ronckers, C. M. and Oldenburger, F. and Balgobind, B. V. and Bosman, P. A. N. and Bel, A. and Alderliesten, T.},
  journal={Med. Phys.},
  volume={45},
  number={6},
  pages={2628--2638},
  year={2018},
  publisher={Wiley Online Library}
}

@article{wang2019how,
	author={Z. Wang and B. Balgobind and M. Virgolin and I. W. E. M. van Dijk and J. Wiersma and C. M. Ronckers and P. A. N. Bosman and A. Bel and T. Alderliesten},
	title={How do patient characteristics and anatomical features correlate to accuracy of organ dose reconstruction for {W}ilms' tumor radiation treatment plans when using a surrogate patient's {CT} scan?},
    journal={J. Radiol. Prot.},
    volume={39},
    number={2},
    pages = {598--619},
    year={2019},
    publisher={IOP Publishing}
}

@article{de2001organ,
  title={Organ weight in 684 adult autopsies: new tables for a {C}aucasoid population},
  author={de la Grandmaison, G. L. and Clairand, I. and Durigon, M.},
  journal={Forensic Sci. Int.},
  volume={119},
  number={2},
  pages={149--154},
  year={2001},
  publisher={Elsevier}
}

@article{SIOPprotocol,
  title={Position paper: rationale for the treatment of {W}ilms tumour in the {UMBRELLA} {SIOP}--{RTSG} 2016 protocol},
  author={van den Heuvel-Eibrink, Marry M and Hol, Janna A and Pritchard-Jones, Kathy and van Tinteren, Harm and Furtw{\"a}ngler, Rhoikos and Verschuur, Arnauld C and Vujanic, Gordan M and Leuschner, Ivo and Brok, Jesper and R{\"u}be, Christian and others},
  journal={Nat. Rev. Urol.},
  volume={14},
  number={12},
  pages={743--752},
  year={2017},
  publisher={Nature Publishing Group}
}

@article{geyer2014uf,
  	year = 2014,
	month = {aug},
	publisher = {{IOP} Publishing},
	volume = {59},
	number = {18},
	pages = {5225--5242},
	author = {Amy M Geyer and Shannon O'Reilly and Choonsik Lee and Daniel J Long and Wesley E Bolch},
	title = {The {UF}/{NCI} family of hybrid computational phantoms representing the current {US} population of male and female children, adolescents, and adults---application to {CT} dosimetry},
	journal = {Phys. Med. Biol.}
}

@article{cassola2011standing,
  year = 2011,
	month = {may},
	publisher = {{IOP} Publishing},
	volume = {56},
	number = {13},
	pages = {3749--3772},
	author = {V F Cassola and F M Milian and R Kramer and C A B de Oliveira Lira and H J Khoury},
	title = {Standing adult human phantoms based on 10th, 50th and 90th mass and height percentiles of male and female {C}aucasian populations},
	journal = {Phys. Med. Biol.}
}

@article{segars2013XCATlibrary,
  title={Population of anatomically variable {4D} {XCAT} adult phantoms for imaging research and optimization},
  author={Segars, W P and Bond, Jason and Frush, Jack and Hon, Sylvia and Eckersley, Chris and Williams, Cameron H and Feng, Jianqiao and Tward, Daniel J and Ratnanather, J T and Miller, M I and others},
  journal={Med. Phys.},
  volume={40},
  number={4},
  pages={043701},
  year={2013},
  publisher={Wiley Online Library}
}

@article{mishra2013evaluation,
 year = 2013,
	month = {jan},
	publisher = {{IOP} Publishing},
	volume = {58},
	number = {4},
	pages = {841--858},
	author = {Pankaj Mishra and Ruijiang Li and Sara St James and Raymond H Mak and Christopher L Williams and Yong Yue and Ross I Berbeco and John H Lewis},
	title = {Evaluation of 3{D} fluoroscopic image generation from a single planar treatment image on patient data with a modified {XCAT} phantom},
	journal = {Phys. Med. Biol.}
}

@book{bishop2006pattern,
  title={Pattern recognition and machine learning},
  author={Bishop, Christopher M},
  year={2006},
  publisher={Springer}
}

@article{virgolin2018feasibility,
  title={On the feasibility of automatically selecting similar patients in highly individualized radiotherapy dose reconstruction for historic data of pediatric cancer survivors},
  author={Virgolin, Marco and van Dijk, I. W. E. M. and Wiersma, Jan and Ronckers, C{\'e}cile M and Witteveen, Cees and Bel, Arjan and Alderliesten, Tanja and Bosman, Peter A. N.},
  journal={Med. Phys.},
  volume={45},
  number={4},
  pages={1504--1517},
  year={2018},
  publisher={Wiley Online Library}
}

@inproceedings{virgolin2017scalable,
 author = {Virgolin, Marco and Alderliesten, Tanja and Witteveen, Cees and Bosman, Peter A. N.},
 title = {Scalable Genetic Programming by Gene-pool Optimal Mixing and Input-space Entropy-based Building-block Learning},
 booktitle = {Proc. Genetic and Evolutionary Computation Conference},
 series = {GECCO '17},
 year = {2017},
 location = {Berlin, Germany},
 pages = {1041--1048},
 numpages = {8},
 publisher = {ACM},
 address = {New York, NY, USA},
}

@article{virgolin2019model,
author = {Virgolin, Marco and Alderliesten, Tanja and Witteveen, Cees and Bosman, Peter A N},
title = {Improving Model-Based Genetic Programming for Symbolic Regression of Small Expressions},
journal = {Evol. Comput.},
volume = {TBA},
number = {TBA},
pages = {TBA (recently accepted)},
year = {2020},
}

@inproceedings{virgolin2018symbolic,
  title={Symbolic regression and feature construction with {GP-GOMEA} applied to radiotherapy dose reconstruction of childhood cancer survivors},
  author={Virgolin, Marco and Alderliesten, Tanja and Bel, Arjan and Witteveen, Cees and Bosman, Peter A N},
  booktitle={Proc. Genetic and Evolutionary Computation Conference},
  series = {GECCO '18},
  pages={1395--1402},
  year={2018},
  organization={ACM},
  address = {New York, NY, USA},
}

@article{virgolin2019machine,
  author = {Marco Virgolin and Ziyuan Wang and Tanja Alderliesten and Peter A. N. Bosman},
title = {{Machine learning for the prediction of pseudorealistic pediatric abdominal phantoms for radiation dose reconstruction}},
volume = {7},
journal = {J. Med. Imag.},
number = {4},
publisher = {SPIE},
pages = {1 -- 25},
year = {2020}
}

@article{virgolin2020spie,
 title={Machine learning for automatic construction of pediatric abdominal phantoms for radiation dose reconstruction},
   author={Virgolin, Marco and Wang, Ziyuan and Alderliesten, Tanja and Bosman, Peter A. N.},
  journal={Proc. SPIE},
  volume={11318},
  pages={1131815},
  year={2020}
}

@article{wang2020spiej,
title={Automatic generation of three-dimensional dose reconstruction data for two-dimensional radiotherapy plans for historically treated patients},
author={Wang, Ziyuan and Virgolin, Marco and Bosman, Peter A N and Crama, Koen and Balgobind, Brian V and Bel, Arjan and Alderliesten, Tanja},
journal={J. Med. Imaging},
volume={7},
number={1},
pages={015001},
year={2020},
publisher={International Society for Optics and Photonics}
}

@article{huijskens2015quantification,
  title={Quantification of renal and diaphragmatic interfractional motion in pediatric image-guided radiation therapy: a multicenter study},
  author={Huijskens, Sophie C and van Dijk, Irma W. E. M. and de Jong, Rianne and Visser, Jorrit and Fajardo, Raquel D{\'a}vila and Ronckers, C{\'e}cile M and Janssens, Geert ORJ and Maduro, John H and Rasch, Coen RN and Alderliesten, Tanja and others},
  journal={Radiother. Oncol.},
  volume={117},
  number={3},
  pages={425--431},
  year={2015},
  publisher={Elsevier}
}

@article{krieger2005monte,
  title={Monte Carlo-versus pencil-beam-/collapsed-cone-dose calculation in a heterogeneous multi-layer phantom},
  author={Krieger, Thomas and Sauer, Otto A},
  journal={Phys. Med. Biol.},
  volume={50},
  number={5},
  pages={859},
  year={2005},
  publisher={IOP Publishing}
}

@article{guidotti2018survey,
  title={A survey of methods for explaining black box models},
  author={Guidotti, Riccardo and Monreale, Anna and Ruggieri, Salvatore and Turini, Franco and Giannotti, Fosca and Pedreschi, Dino},
  journal={ACM Comput. Surv.},
  volume={51},
  number={5},
  pages={1--42},
  year={2018},
  publisher={ACM}
}

@article{adadi2018peeking,
  title={Peeking inside the black-box: {A} survey on {E}xplainable {A}rtificial {I}ntelligence ({XAI})},
  author={Adadi, Amina and Berrada, Mohammed},
  journal={IEEE Access},
  volume={6},
  pages={52138--52160},
  year={2018},
  publisher={IEEE}
}

\end{document}